**Type of manuscript**: Practice of Epidemiology

**Manuscript title**: Examining the Association between Estimated Prevalence and Diagnostic Test Accuracy Using Directed Acyclic Graphs

**Authors**:

Yang Lu[a]; Robert W. Platt[a,b]; Nandini Dendukuri[a,c]

[a]Department of Epidemiology, Biostatistics and Occupational Health, McGill University, Montreal, QC, Canada

[b]Department of Pediatrics, McGill University, Montreal, QC, Canada

[c]Department of Medicine, McGill University, Montreal, QC, Canada

**Corresponding Author**:

Nandini Dendukuri,

5252 boul. de Maisonneuve Bureau 3F.50

Montreal, Quebec, H4A 3S5

(514) 934-1934 ext: 36916

nandini.dendukuri@mcgill.ca

**Running Head**: ASSOCIATION BETWEEN PREVALENCE AND TEST ACCURACY

**Sources of Financial support**: Natural Sciences and Engineering Research Council (NSERC) Grant number RGPIN/06713-2019

**Conflict of interest**: The authors have no conflicts of interest to disclose.

**Reproducibility**: The R scripts used for all simulation studies are available on request.

# Examining the Association between Estimated Prevalence and Diagnostic Test Accuracy using Directed Acyclic Graphs

## Abstract

There have been reports of correlation between estimates of prevalence and test accuracy across studies included in diagnostic meta-analyses. It has been hypothesized that this unexpected association arises because of certain biases commonly found in diagnostic accuracy studies. A theoretical explanation has not been studied systematically. In this work, we introduce directed acyclic graphs to illustrate common structures of bias in diagnostic test accuracy studies and to define the resulting data-generating mechanism behind a diagnostic meta-analysis. Using simulation studies, we examine how these common biases can produce a correlation between estimates of prevalence and index test accuracy and what factors influence its magnitude and direction. We found that an association arises either in the absence of a perfect reference test or in the presence of a covariate that simultaneously causes spectrum effect and is associated with the prevalence (confounding). We also show that the association between prevalence and accuracy can be removed by appropriate statistical methods. In the risk of bias evaluation in diagnostic meta-analyses, an observed association between estimates of prevalence and accuracy should be explored to understand its source and to adjust for latent or observed variables if possible.

## Introduction

Diagnostic test accuracy (DTA) studies are an important early step in diagnostic test evaluation. These studies estimate an index test's ability to detect the target condition in terms of its sensitivity and specificity (i.e., sensitivity = P(Index test is positive | Target condition is present); specificity = P(Index test is negative | Target condition is absent)). As sensitivity and specificity involve conditioning on the target condition, by their very definition, these parameters should be independent of the prevalence of the target condition (i.e., P(Target condition is positive)). Yet, a growing number of empirical studies, based on meta-analyses of DTA studies, have noted the contrary.[1-4] In some cases, the association is possibly artefactual, while in others it may be real.

Leeflang et al. drew attention to the fact that the direction of the observed correlation is not always the same [3,4] and hypothesized it may be the result of well-recognized biases in the design of individual DTA studies. DTA studies typically use a non-randomized design and are therefore prone to well-recognized biases including reference standard error bias, partial verification bias, spectrum effect, confounding and the misassumption of conditional independence [5-10]. For example, it is well known that individual DTA studies relying on an imperfect reference standard can lead to bias in estimates of both diagnostic test accuracy and target condition prevalence (i.e., reference standard error bias).[4,11-14] This may explain how a spurious association between prevalence and accuracy arises across settings. Other work has documented that the performance of a diagnostic test may differ considerably across patient populations or clinical settings (i.e., spectrum effect).[1,4,10,15-19] If the prevalence also varies across these settings, it may lead to a biased spurious association between prevalence and accuracy across settings (i.e., confounding). It is also known that the presence of partial verification bias can result in overestimation in both sensitivity and prevalence, and an underestimation of specificity.[3,4,9] This can potentially create a spurious relationship between prevalence and accuracy.

In etiologic studies, directed acyclic graphs (DAGs) are widely used to depict the causal relations between variables, and to identify the presence of various biases (e.g. confounding, misclassification and selection bias) that could impact identification of these relations.[20-26] DAGs are also useful for illustrating data generating processes and missing data mechanisms.[27] However, DAGs have not been used in the context of DTA studies.[28]

Our primary objective is to confirm how common biases in individual DTA studies influence the magnitude and direction of the association between estimates of target condition prevalence and test accuracy across studies in a meta-analysis setting. We achieve this by employing DAGs to illustrate the biases and to support simulation studies. Finally, we attempt to correct the association through statistical adjustments.

## Methods

### DAGs for illustrating Structure of Bias in DTA studies

A DAG illustrates causal relationships using two components: vertices and edges. In the context of clinical and epidemiological research, the vertices are variables such as exposure/treatment, outcome and confounder variables. DAGs may include both observed and latent variables. The edges are arrows from the cause (e.g. exposure variable) to the result (e.g. outcome variable). A characteristic feature of a DAG is that it has no closed cycles as the outcome cannot cause the exposure (more broadly, a variable cannot cause another variable that precedes it temporally). DAGs may include both observed and latent variables. To create a DAG, we begin by enumerating the relevant variables or vertices and then depict any causal relations between them using edges.

In the context of DTA studies, the variables of interest are: i) the target condition, ii) one or more index test results, and iii) covariates or selection criteria that affect the target condition and/or the index test. The association of interest is the relationship between the target condition and index test result(s) (i.e., sensitivity and specificity). When a gold standard or perfect reference test is not available, we consider the target condition is latent and assume that the reference test result is a distinct variable from the target condition. We use a shaded circle to represent a latent (unmeasured) variable and a square to represent conditioning on a variable. Additional observed or latent variables may be involved which result in the presence of different structures of bias. The examples in the following section illustrate this.

Table 1 lists DAGs that illustrate the structure of reference standard error bias, spectrum effect, confounding, partial verification bias, and bias due to misassumption of conditional independence. A real-life example is provided in each case. These graphical representations serve three purposes. First, they illustrate the structure of the bias. Second, they provide a systematic framework for designing appropriate simulation settings and data generating mechanisms that can accurately reflect the causal relationships in DTA studies. Third, they can inform strategies for bias correction.

It is interesting to note that reference standard bias is counterpart to information bias due to misclassification, and verification bias is counterpart to selection bias in the context of etiological research studies.[42]

## Simulation

We now define notation that will allow us to simulate data that arise in a DTA meta-analysis under different scenarios. A more detailed summary of the complete simulation design and rationales, reported according to the ADEMP framework, is available in Supplementary Material S1.

**Notation:** Let us assume that $P$ studies are observed. Let $D_{ij}$ denote the target condition of the $j^{th}$ subject in the $i^{th}$ study ($j = 1, \ldots, N_i$ and $i = 1, \ldots, P$), and $\pi_i$ denote the prevalence of the target condition in the $i^{th}$ study. Let $T_{ijk}$ denote the $k^{th}$ test obtained from the $j^{th}$ subject in the $i^{th}$ study such that $k = 1$ for the reference test and $k = 2$ for the index test. Let $S_{ik}$ denote the sensitivity of the $k^{th}$ test in the $i^{th}$ study and $C_{ik}$ denote the specificity of the $k^{th}$ test in the $i^{th}$ study. Let $R_{ij}$ denote the covariate measured on the $j^{th}$ subject in the $i^{th}$ study which influences the accuracy of the index test thus inducing spectrum effect, and $p_i$ denote the prevalence of this covariate in the $i^{th}$ $study$. Let $V_i$ denote the verification variable/procedure in the $i^{th}$ study.

**Scenarios:** For each bias, we generated data under different test accuracy values that are likely to be observed in practice, e.g. where both tests have sufficiently high specificity and sensitivity values such that the diagnostic value (sensitivity + specificity) exceeds 1. The details regarding the settings for the simulation parameters for each structure of bias are presented in Table 1. In all scenarios, the index test sensitivity and specificity were set around 0.8-0.9, while reference test sensitivity varied more. To isolate the effect of different biases, we allowed the reference test to have perfect accuracy in some scenarios. Reference test sensitivities varied from 0.65-1 and specificities were 0.95-1. Values of the prevalence varied from 0.1 to 0.9 across studies. These values for imperfect tests and prevalence were inspired by examples of diagnostic tests used in pediatric and adult tuberculosis (TB) detection.[43,44]

**Data-generating Mechanism**: We simulated 10,000 studies, each containing 500 subjects. For each subject, we generated data on one target condition, one reference standard result, one index test result, and when applicable, one auxiliary variable (e.g., R). Within each study, index test sensitivity was estimated as the proportion of positive index test results among all positive reference standard results, assuming the reference standard to be perfect. Similarly, specificity was calculated as the proportion of negative index test results among all negative reference standard results. The prevalence of the target condition is estimated as the proportion of subjects who are positive on the reference standard within each study. This procedure generated data at the subject-level, calculated accuracy and prevalence estimates for each of the 10,000 studies.

**Examine Association between Accuracy and Prevalence Estimates:** To visualize and quantify the induced association between index test accuracy and prevalence estimates across studies under different biases in the individual studies, we created scatter plots and calculated Spearman's rank correlation coefficient to capture potential nonlinear relationships [45].

Statistical Adjustment to Remove the Association between Prevalence and Accuracy

We investigated whether the spurious relationship between prevalence and accuracy can be removed using a latent class bivariate meta-analysis model (LCBM) [46,47] and its variants [48] (specifically extended for partial verification bias). For biases involving subgroup effects (spectrum effect, confounding, and conditional independence violations), we applied LCBM using a subgroup analysis strategy. The statistical adjustment was conducted across all bias settings. Due to computational constraints, we limited our investigation to a setting with 100 studies in the meta-analysis, each containing 500 subjects. A detailed model description, MCMC configurations and the corresponding JAGS [49] implementation can be found in Supplementary File S2 and S3. Since we employed Bayesian inference for estimation, we summarized and reported the parameters of interest using posterior quantiles.

Results

Figure 1-5 show the relationships between the estimated sensitivity and specificity of the index test and the estimated prevalence under five different structures of bias, across various reference standard scenarios.

**Impact of reference standard error bias:** In Figure 1A and 1B, no obvious association was observed for either the sensitivity ($\rho = -0.026$) or specificity ($\rho = 0.01$) when the reference test was perfect. On the other hand, in Figure 1A, a positive association ($\rho = 0.870, 0.864, 0.856$) was observed for the settings when the reference test had imperfect accuracy. The magnitude of the association did not vary much across settings for the sensitivity of the reference test, while its specificity was fixed at 0.95. In contrast, in Figure 1B, the associations between the specificity and prevalence were noted to be stronger in magnitude ($\rho = -0.975, -0.968, -0.936$).

**Impact of spectrum effect:** In Figure 2A and 2B, negligible associations were observed for both the sensitivity ($\rho = -0.004$) and specificity ($\rho = -0.003$) when the reference standard has a perfect accuracy. When using imperfect reference standards, sensitivity showed moderate positive correlations ($\rho = 0.645, 0.632, 0.600$) as shown in Figure 2A across different reference test sensitivity values with reference standard specificity fixed at 0.95. In contrast, in Figure 2B, specificity showed progressively stronger negative correlations ($\rho = -0.930, -0.893, -0.786$), with the magnitude increasing as reference test sensitivity decreased.

**Impact of confounding:** In the confounding scenarios illustrated in Figure 3, we found the association between the estimated sensitivity of the index test and the estimated prevalence could be positive or negative even when the reference test is perfect. Figure 3A shows a negative association ($\rho = -0.599$) when the reference standard is perfect, while weak positive associations ($\rho = 0.461, 0.395, 0.335$) were noted across settings when the reference test is imperfect. Figure 3B illustrates a negative association between the estimated specificity of the index test and prevalence when the reference standard is imperfect. The associations with different magnitudes in Figure 3B were observed by varying the sensitivity of the reference test, while its specificity was fixed at 0.95. Lower sensitivity of the reference test corresponded to a stronger association ($\rho = -0.956, -0.947, -0.915, -0.621$).

**Impact of partial verification bias**: Figures 4A and 4B illustrate scenarios that yield results comparable to those observed in the context of reference standard bias alone. A negligeable association ($\rho = 0.04$ and -0.053, respectively) was observed across the simulated studies when the reference standard was free of error. This finding suggests that the spurious relationship between sensitivity and specificity of the index test and prevalence cannot arise from partial verification bias alone when the reference standard is perfect. However, it is evident from both figures that the estimated sensitivity and specificity deviate from their true value, irrespective of reference standard accuracy.

In addition, the association in Figure 4A was slightly weaker in magnitude ($\rho = 0.821, 0.808, 0.806$) when a combination of reference standard error bias and the partial verification bias is present, compared to the reference standard bias alone. Figure 4B reveals that the association between the specificity and prevalence became stronger ($\rho = -0.967, -0.955, -0.918$) as we decreased the sensitivity of the reference standard while keeping the specificity fixed at 0.95, given the presence of partial verification bias. Similar patterns were also observed when varying the verification rate (see Supplementary Material S4).

**Impact of misassumption of conditional independence:** Note that only imperfect reference standard scenarios are presented in Figure 5, as conditional dependence requires imperfect accuracy in both the reference standard and index test. Figure 5A shows similar positive associations ($\rho = 0.704, 0.697, 0.687$) between the sensitivity of index test and prevalence across all imperfect accuracy settings for the reference test. Figure 5B illustrates negative relationships between the estimated specificity of the index test and prevalence. These relationships are again weakened as the true sensitivity of the reference test increases ($\rho - 0.928, -0.896, -0.811$) while its specificity is fixed at 0.95.

**Statistical adjustment of each structure of bias:** We applied the LCBM and its variant to address various forms of bias in the simulated data. These approaches effectively eliminated the associations between index test accuracy parameters and prevalence estimates that are introduced by different structures of bias. Visualizations demonstrating these bias corrections are available in Supplementary Material S5.

# Discussion

In this manuscript, we have systematically investigated how the reported association between estimates of diagnostic test accuracy and the target condition prevalence arises in DTA meta-analyses. Whereas the literature so far referred to the existence of an association between the parameters themselves, we have shown that any association is in fact between the estimates of the parameters.

Through simulation studies we showed that the apparent relationship across studies could be the result of biases that arise in individual studies. We found that reference standard error bias can result in a spurious association between estimates of accuracy and prevalence.

The presence of spectrum effect alone does not seem to cause an association between accuracy and prevalence estimates across studies. However, simply reporting a single estimate when subgroup heterogeneity in test accuracy is present can be misleading and have limited clinical usefulness, and it is advisable to conduct stratified analysis using covariate $R$ as stratification variable.[32]

On the other hand, confounding can result in a spurious association irrespective of whether the reference standard is perfect or imperfect. In particular, we notice a negative association between sensitivity and prevalence when a perfect reference standard is used. This negative association can be explained by the simulation setup where the confounding variable $R$ was set to be associated with a low sensitivity of index test and high risk of having the target condition. In the presence of confounding, it is probably also advisable to report results within strata defined by the covariate $R$ or through covariate adjustment.[50] If we fail to adjust for the common cause that affect the target condition and index test accuracy, the estimated accuracy will be confounded and biased. Pooling results across studies would mask the actual value of the index test accuracy in practice in different settings.

The presence of partial verification bias alone does not seem to cause an association between accuracy and prevalence across studies. Still, the presence of partial verification bias led to biased estimation of test accuracy and target condition prevalence.

We have also shown how DAGs can be helpful tools in illustrating the structure of common biases in the context of DTA studies, though our DAG for each structure represents just one of potentially many ways to conceptualize these relationships. We recommend that future DTA studies or their meta-analyses include a corresponding DAG to enhance clarity and facilitate understanding regarding any biases that may be present in individual studies. Besides, we note that confounding and misassumption of conditional independence require spectrum effect in the index test accuracy, which is reflected in their corresponding probabilistic definition.

The application of LCBM and subgroup analyses showed that the biased relationship induced by common biases can be successfully removed. However, it should be noted that in practice, multiple biases could occur simultaneously. Additional adjustments to the LCBM would be needed in that case.

A possible drawback of our work is that the simulation settings are not exhaustive. In practice, the range of test accuracy and prevalence may be greater. Also, in practice, more variables and multiple sources of bias may be simultaneously involved. Nonetheless, it is helpful to examine each bias in isolation to understand its consequence.

# Conclusion

A spurious relationship between estimates of prevalence and test accuracy can arise when the same covariate affects both of index test accuracy and target condition prevalence irrespective of whether a perfect or imperfect reference standard is used. The presence of reference standard error bias results in a spurious relationship between estimates of test accuracy and prevalence. This spurious relationship can be removed by modeling the accuracy of the reference test through latent class meta-analysis models. Therefore, the presence of an association between prevalence and test accuracy is a cue for researchers to adjust the DTA meta-analysis appropriately. Whereas researchers are conscious of reference standard bias and spectrum effect and may even routinely document them using tools like QUADAS [6], latent class models to adjust for reference standard bias or stratified analyses to adjust for spectrum effect are rarely applied. More efforts are needed to encourage usage of these established methods and to make them more accessible.

Table 1. DAGs and Simulation Settings for Each Structure of Biases. The numerical values of the parameters in the data-generating mechanism cover broader ranges and are not intended to represent the example(s) provided under each bias. $D$ denotes the target condition; $R$ denotes the risk factor/common cause; $T_1$ denotes the reference standard; $T_2$ denotes the index test; $V$ denotes the verification indicator.

| Structure of Bias | DAGs | Data-generation Parameters |
|---|---|---|
| **Reference Standard Error Bias:** It occurs when the reference standard ($T_1$) has an intrinsic imperfect sensitivity and/or specificity and thus may provide an erroneous measurement of the target condition [11,13]. <br><br> **Example:** When estimating the accuracy of GeneXpert ($T_2$) for pediatric pulmonary TB by comparing it to cell culture (an imperfect reference standard), the sensitivity and specificity of GeneXpert may be under- or overestimated [29]. | 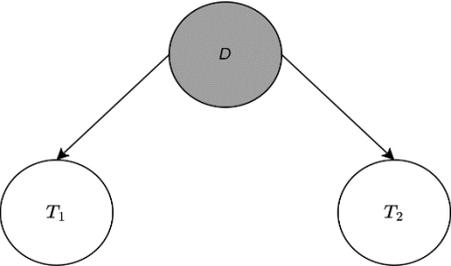 | $D_{ij}\|\pi_i \sim Bern(\pi_i)$ <br> $\pi_i \sim Unif(0.1, 0.9)$ <br> $T_{ijk}\|D_{ij}=1 \sim Bern(S_k)$ <br> $T_{ijk}\|D_{ij}=0 \sim Bern(1-C_k)$ <br> $S_1 = 0.7, 0.8, 0.9, 1$ <br> $C_1 = 0.95, 1$ <br> $S_2 = 0.9$ <br> $C_2 = 0.9$ <br> $i = 1, 2, \ldots, 10000$ <br> $j = 1, 2, \ldots, 500$ <br> $k = 1, 2$ |
| **Spectrum Effect:** It occurs when the index test ($T_2$) performs differently in different groups of subjects defined by one or more covariates ($R$).[10,16,30] The additional variable, $R$, may be observed or latent. <br><br> **Example:** Enzyme Immunoassay for Chlamydia trachomatis has higher sensitivity and specificity in younger patient than older patient ($R$: Age > 24 or Age ≤ 24). In contrast, when stratifying by clinic type ($R$: Family planning or Sexually transmitted disease), enzyme immunoassay | 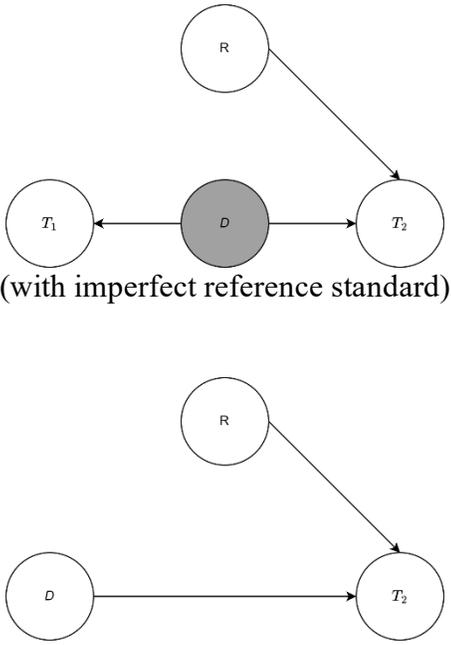 <br> (with imperfect reference standard) <br><br> (with perfect reference standard) | $D_{ij}\|\pi_i \sim Bern(\pi_i)$ <br> $\pi_i \sim Unif(0.1, 0.9)$ <br> $T_{ijk}\|D_{ij}=1, R_{ij}=1 \sim Bern(S_{k1})$ <br> $T_{ijk}\|D_{ij}=1, R_{ij}=0 \sim Bern(S_{k0})$ <br> $T_{ijk}\|D_{ij}=0, R_{ij}=1 \sim Bern(1-C_{k1})$ <br> $T_{ijk}\|D_{ij}=0, R_{ij}=0 \sim Bern(1-C_{k0})$ <br> $R_{ij} \sim Bern(p_i)$ <br> $p_i \sim Beta(1,1)$ <br> $S_{11} = S_{10} = 0.7, 0.8, 0.9, 1$ <br> $C_{11} = C_{10} = 0.95, 1$ <br> $S_{21} = 0.8$ <br> $S_{20} = 0.9$ <br> $C_{21} = 0.8$ <br> $C_{20} = 0.9$ <br> $i = 1, 2, \ldots, 10000$ <br> $j = 1, 2, \ldots, 500$ <br> $k = 1, 2$ |
| has a higher specificity in sexually transmitted disease strata.[31,32] | | |
| **Confounding:** It occurs when both target condition and index test performance are dependent on an additional variable $R$. The additional variable, $R$, may be observed or latent.<br><br>**Example:** Human Immunodeficiency Virus (HIV) lowers the immune response of children with TB, thus lowering the accuracy of the tuberculin skin test (TST) which detects immune response.[29] HIV also increases the risk a child becomes infected with TB. Thus, as prevalence of HIV increases across studies, sensitivity of TST is expected to decrease. | 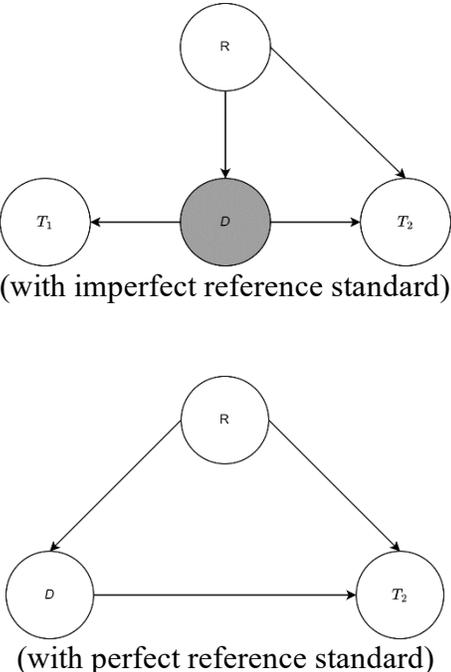<br>(with imperfect reference standard)<br><br>(with perfect reference standard) | $D_{ij}\|\pi_i \sim Bern(\pi_i)$<br>$\pi_i\|R_{ij}=1 \sim Unif(0.7,0.9)$<br>$\pi_i\|R_{ij}=0 \sim Unif(0.1,0.3)$<br>$R_{ij} \sim Bern(p_i)$<br>$p_i \sim Beta(1,1)$<br>$T_{ijk}\|D_{ij}=1, R_{ij}=1 \sim Bern(S_{k1})$<br>$T_{ijk}\|D_{ij}=1, R_{ij}=0 \sim Bern(S_{k0})$<br>$T_{ijk}\|D_{ij}=0, R_{ij}=1 \sim Bern(1-C_{k1})$<br>$T_{ijk}\|D_{ij}=0, R_{ij}=0 \sim Bern(1-C_{k0})$<br>$S_{11}=S_{10}=0.7, 0.8, 0.9, 1$<br>$C_{11}=C_{10}=0.95, 1$<br>$S_{21}=0.8$<br>$S_{20}=0.9$<br>$C_{21}=0.8$<br>$C_{20}=0.9$<br>$i=1,2,\ldots,10000$<br>$j=1,2,\ldots,500$<br>$k=1,2$ |
| **Partial Verification Bias (work-up bias):** It occurs when subjects' likelihood of being tested (V=1) with the reference standard ($T_1$) (i.e., their likelihood of having complete data) is determined by their result on the initial index test ($T_2$).[12,16,33,34] | 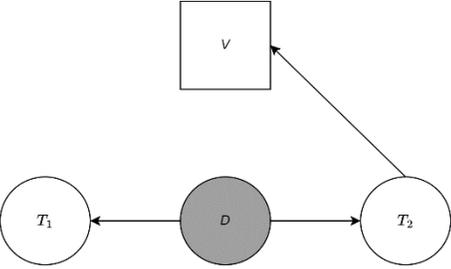<br>(with imperfect reference standard) | $D_{ij}\|\pi_i \sim Bern(\pi_i)$<br>$\pi_i \sim Unif(0.1,0.9)$<br>$V_{ij}=1\|T_{ij2}=1 \sim Bern(1)$<br>$V_{ij}=1\|T_{ij2}=0, pv_i \sim Bern(pv_i)$<br>$pv_i \sim Uniform(0.5,0.9)$<br>$T_{ijk}\|D_{ij}=1 \sim Bern(S_k)$<br>$T_{ijk}\|D_{ij}=0 \sim Bern(1-C_k)$<br>$S_1 = 0.7, 0.8, 0.9, 1$<br>$C_1 = 0.95, 1$<br>$S_2 = 0.9$ |
| Recent methodological work has shown that selection bias can arise through various mechanisms beyond conditioning on colliders, and partial verification bias represents one such example that does not require conditioning on a collider.[21,35,36]<br><br>**Example:** To evaluate the accuracy of polymerase chain reaction (PCR) for detecting human pappiloma virus (HPV), all subjects who tested positive on PCR received the invasive reference test (colposcopy), but only a random sample of those who were negative on PCR received it.[37] | 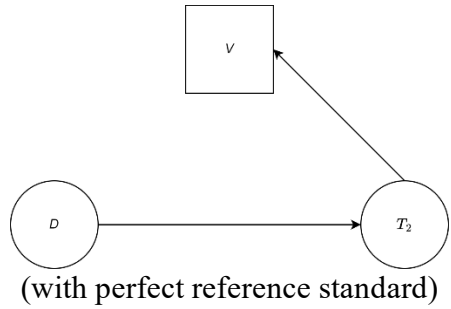<br>(with perfect reference standard) | $C_2 = 0.9$<br>$i = 1,2,...,10000$<br>$j = 1,2,...,500$<br>$k = 1,2$ |
| **Bias due to Misassumption of Conditional Independence:**<br>It occurs when both index test and reference standard are dependent on an additional variable ($R$) besides the target condition ($D$), and when R is ignored in the analysis.[38-41] The additional variable, $R$, may be observed or latent.<br><br>**Example:** Among children with pediatric pulmonary TB, the results of GeneXpert and culture may be correlated due to the dependence of both tests on the bacterial load in the sample.[29] Thus, the accuracy of GeneXpert with respect | 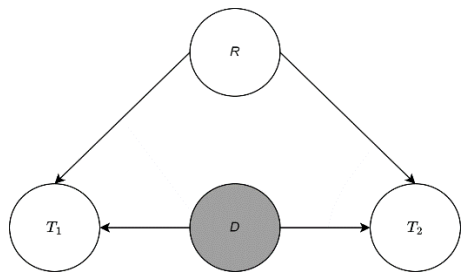 | $D_{ij}\|\pi_i \sim Bern(\pi_i)$<br>$\pi_i \sim Unif(0.1,0.9)$<br>$R_{ij} \sim Bern(p_i)$<br>$p_i \sim Beta(1,1)$<br>$T_{ijk}\|D_{ij}=1, R_{ij}=1 \sim Bern(S_{k1})$<br>$T_{ijk}\|D_{ij}=0, R_{ij}=1 \sim Bern(1-C_{k1})$<br>$T_{ijk}\|D_{ij}=1, R_{ij}=0 \sim Bern(S_{k0})$<br>$T_{ijk}\|D_{ij}=0, R_{ij}=0 \sim Bern(1-C_{k0})$<br>$S_{11} = 0.6, 0.7, 0.8$<br>$C_{11} = 0.85$<br>$S_{10} = 0.7, 0.8, 0.9$<br>$C_{10} = 0.95$<br>$S_{21} = 0.8$<br>$C_{21} = 0.8$<br>$S_{20} = 0.9$<br>$C_{20} = 0.9$<br>$i = 1,2,...,10000$<br>$j = 1,2,...,500$<br>$k = 1,2$ |

| to culture, its accuracy will be overestimated. | | |

FIGURE CAPTIONS

Figure 1. Relationship between estimated sensitivity (1A) and specificity (1B) of index test ($T_2$) and estimated prevalence under reference standard error bias with varying reference standard accuracies. Index test parameters were fixed (sensitivity=0.9, specificity=0.9). Four reference standard scenarios are shown: Setup 1 (red circles): sensitivity=0.7, specificity=0.95; Setup 2 (green triangles): sensitivity=0.8, specificity=0.95; Setup 3 (blue squares): sensitivity=0.9, specificity=0.95; Setup 4 (purple crosses): perfect reference standard (sensitivity=1.0, specificity=1.0). Each point represents one simulated study (n=500 subjects per study).

Figure 2. Relationship between estimated sensitivity (2A) and specificity (2B) of index test ($T_2$) and estimated prevalence under spectrum effect with varying reference standard accuracies. Index test parameters varied by subgroup R (R+: sensitivity=0.80, specificity=0.80; R-: sensitivity=0.90, specificity=0.90). Four reference standard scenarios are shown: Setup 1 (red circles): sensitivity=0.70, specificity=0.95; Setup 2 (green triangles): sensitivity=0.80, specificity=0.95; Setup 3 (blue squares): sensitivity=0.90, specificity=0.95; Setup 4 (purple crosses): perfect reference standard (sensitivity=1.00, specificity=1.00). Each point represents one simulated study (n=500 subjects per study).

Figure 3. Relationship between estimated sensitivity (3A) and specificity (3B) of index test ($T_2$) and estimated prevalence under confounding with varying reference standard accuracies. Index test parameters varied by subgroup R (R+: sensitivity=0.80, specificity=0.80; R-: sensitivity=0.90, specificity=0.90). Prevalence varied by subgroup R (R+: ranging from 0.7 to 0.9; R-: ranging from 0.1 to 0.3). Four reference standard scenarios are shown: Setup 1 (red circles): sensitivity=0.70, specificity=0.95; Setup 2 (green triangles): sensitivity=0.80, specificity=0.95; Setup 3 (blue squares): sensitivity=0.90, specificity=0.95; Setup 4 (purple crosses): perfect reference standard (sensitivity=1.00, specificity=1.00). Each point represents one simulated study (n=500 subjects per study).

Figure 4. Relationship between estimated sensitivity (4A) and specificity (4B) of index test ($T_2$) and estimated prevalence under partial verification bias with varying reference standard accuracies. Index test parameters were fixed (sensitivity=0.9, specificity=0.9). Verification rate was 100% for index test positives and varied uniformly between 0.5 and 0.9 for index test negatives. Four reference standard scenarios are shown: Setup 1 (red circles): sensitivity=0.7, specificity=0.95; Setup 2 (green triangles): sensitivity=0.8, specificity=0.95; Setup 3 (blue squares): sensitivity=0.9, specificity=0.95; Setup 4 (purple crosses): perfect reference standard

(sensitivity=1.0, specificity=1.0). Each point represents one simulated study (n=500 subjects per study).

Figure 5. Relationship between estimated sensitivity (5A) and specificity (5B) of index test ($T_2$) and estimated prevalence under misassumption of conditional independence with varying reference standard accuracies. Index test parameters varied by subgroup R (R+: sensitivity=0.8, specificity=0.8; R-: sensitivity=0.9, specificity=0.9). Three reference standard scenarios are shown: Setup 1 (red circles): R+ sensitivity=0.6, specificity=0.85, R- sensitivity=0.7, specificity=0.95; Setup 2 (green triangles): R+ sensitivity=0.7, specificity=0.85, R- sensitivity=0.8, specificity=0.95; Setup 3 (blue squares): R+ sensitivity=0.8, specificity=0.85, R- sensitivity=0.9, specificity=0.95. Each point represents one simulated study (n=500 subjects per study).

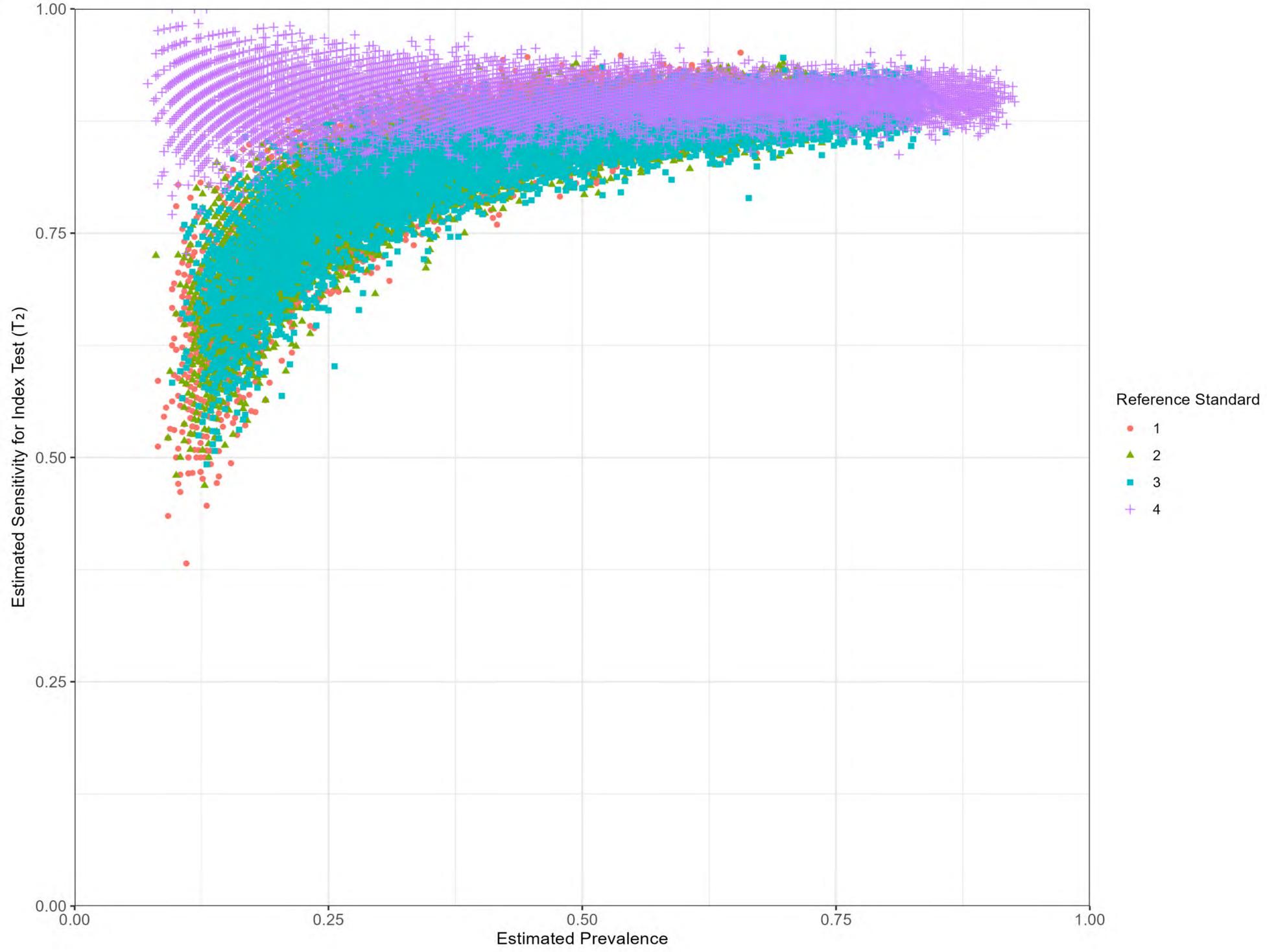

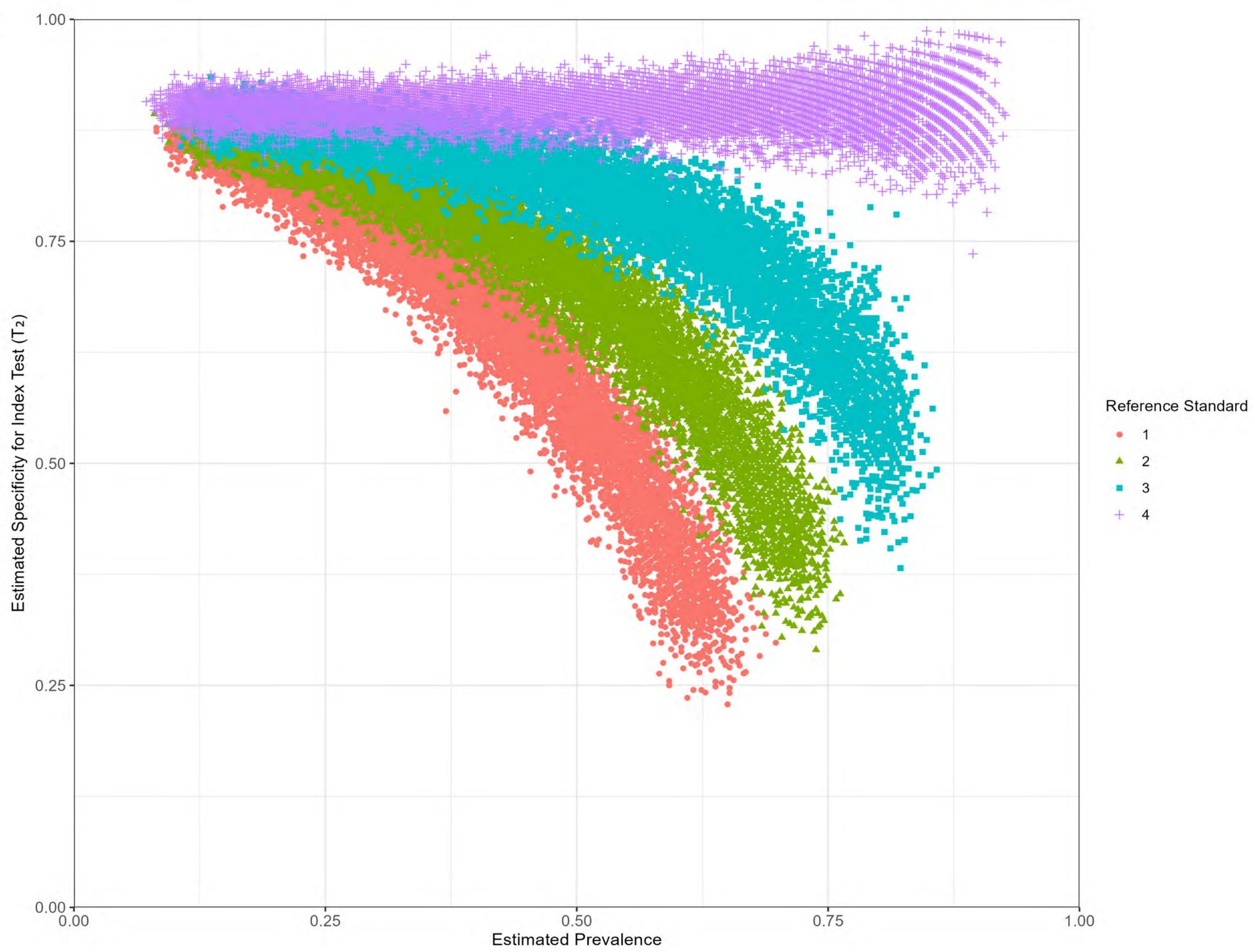

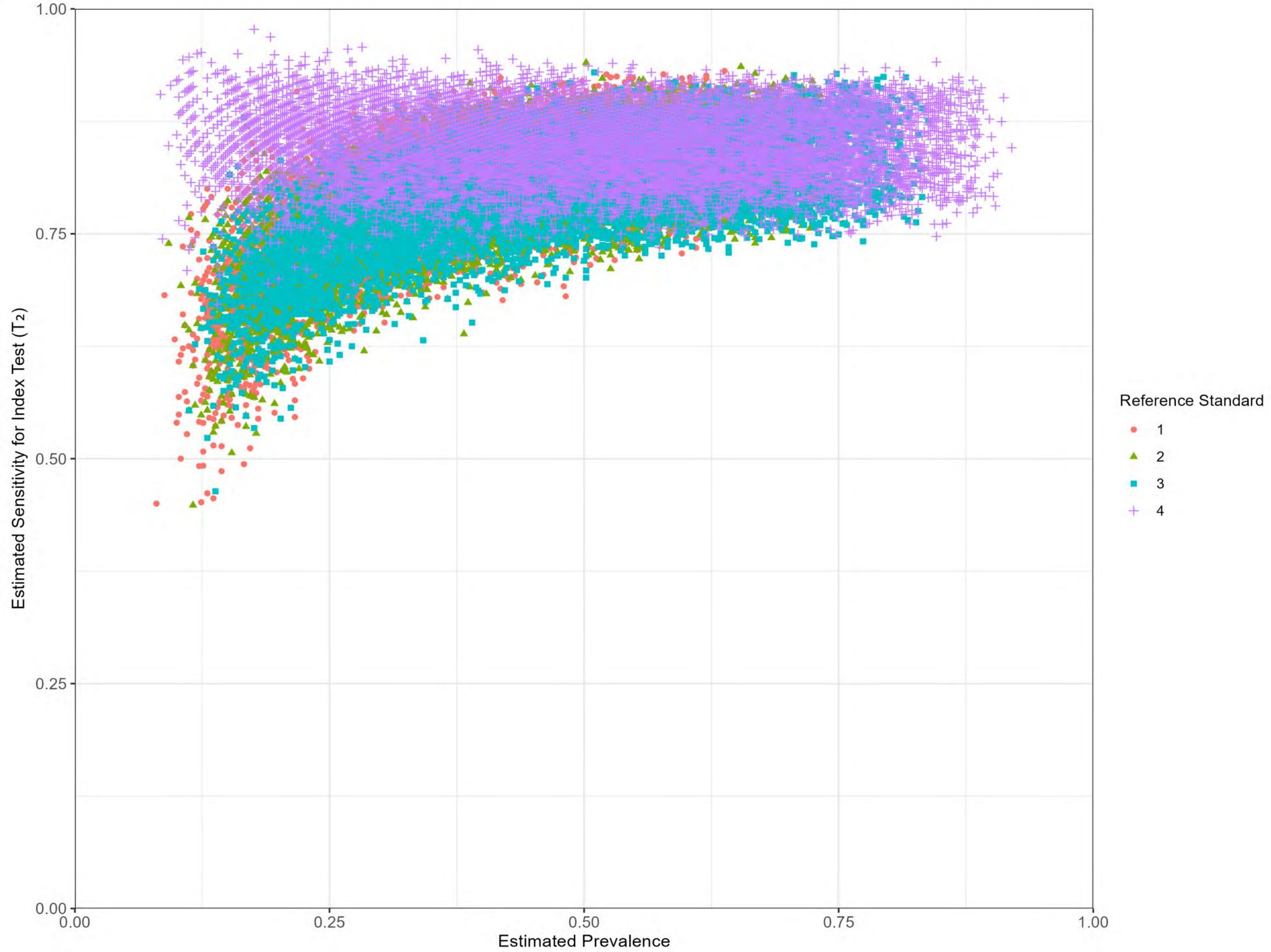
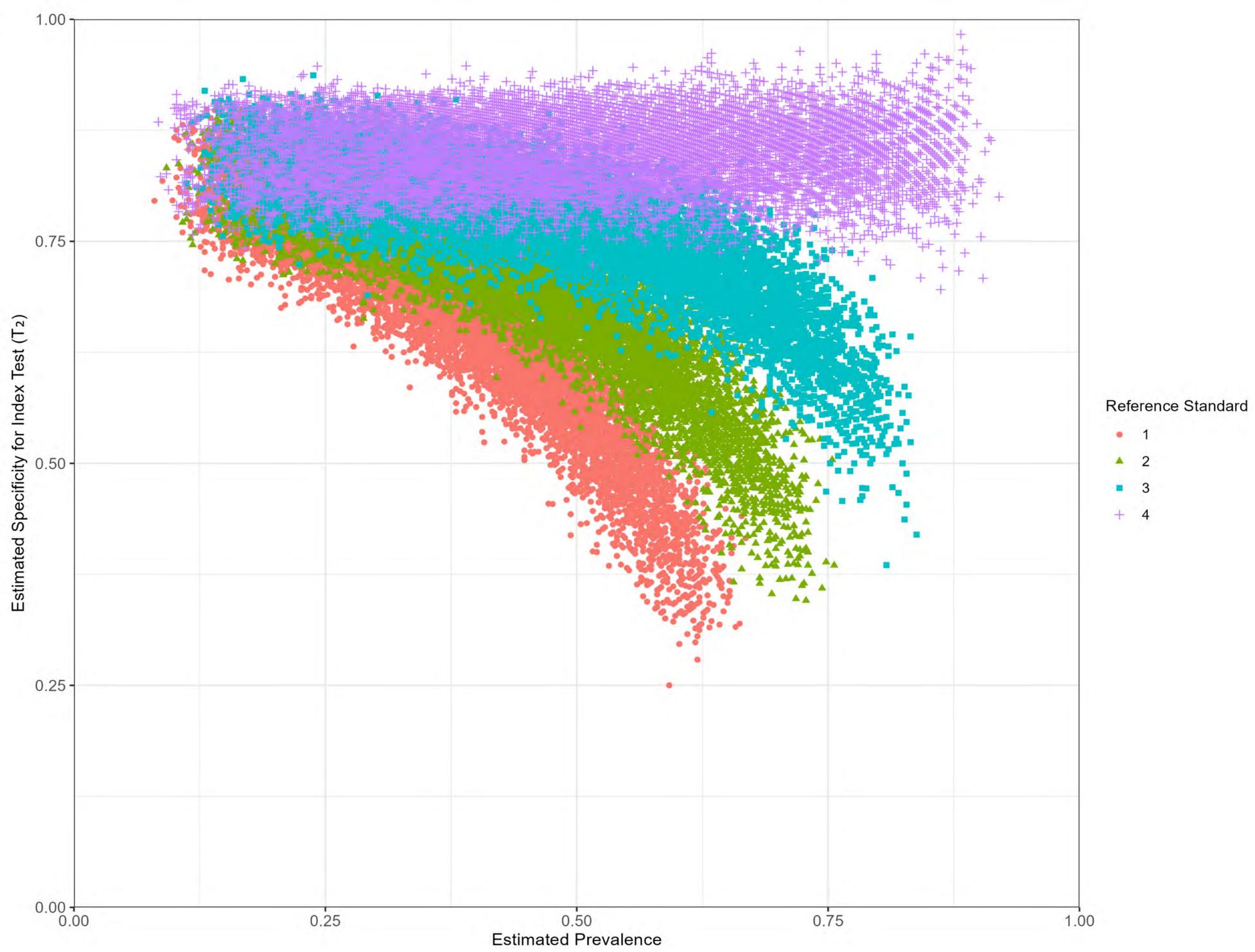

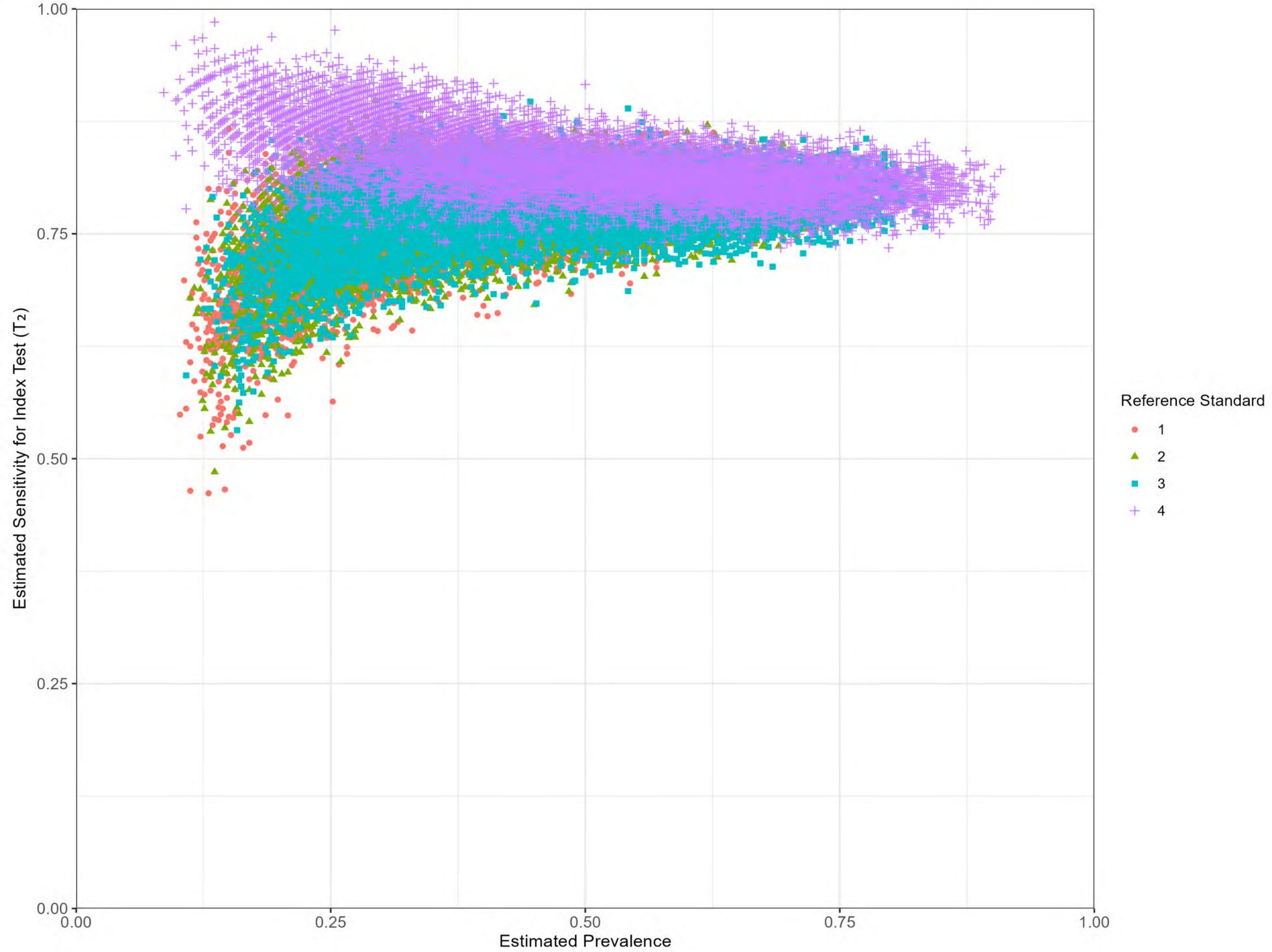
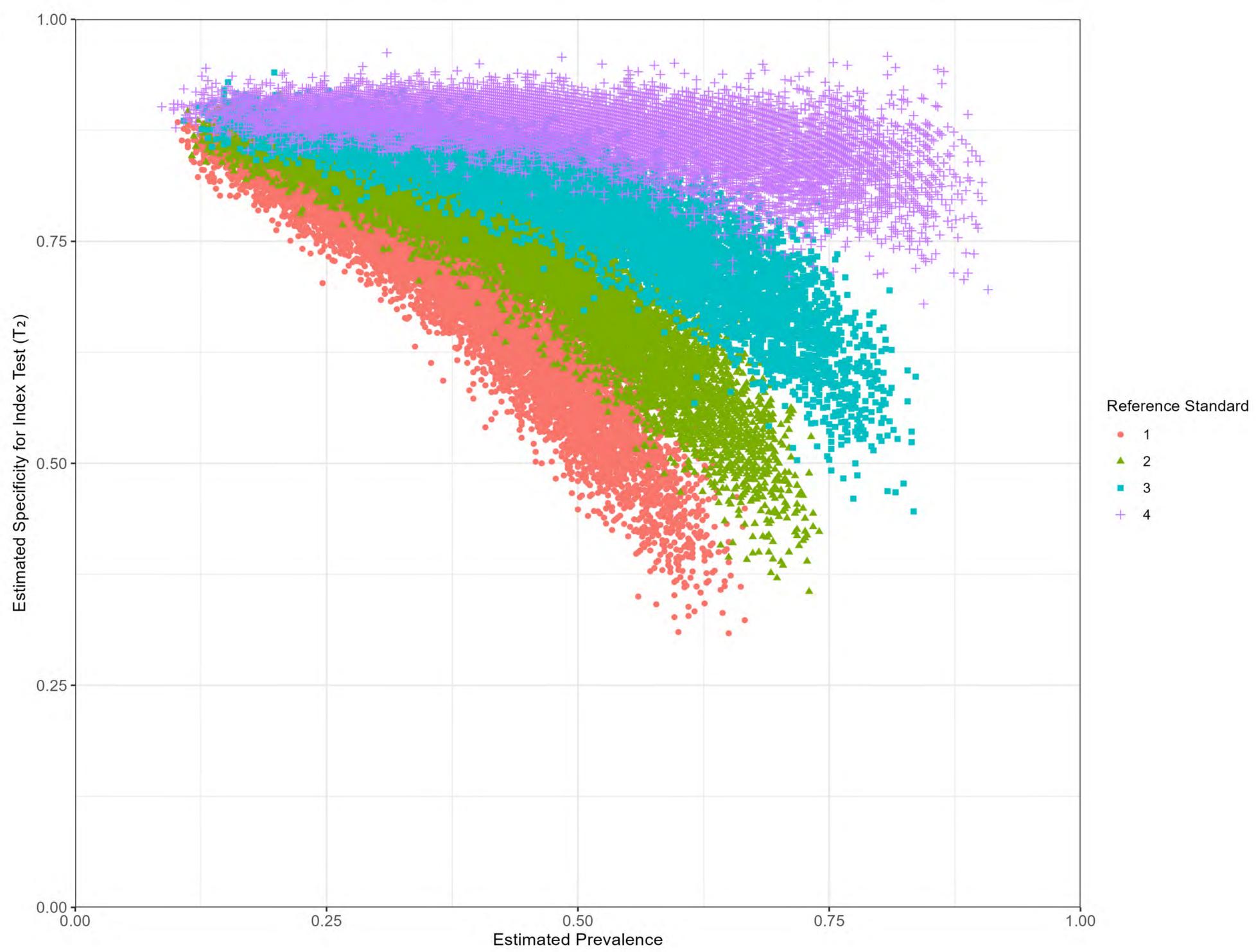

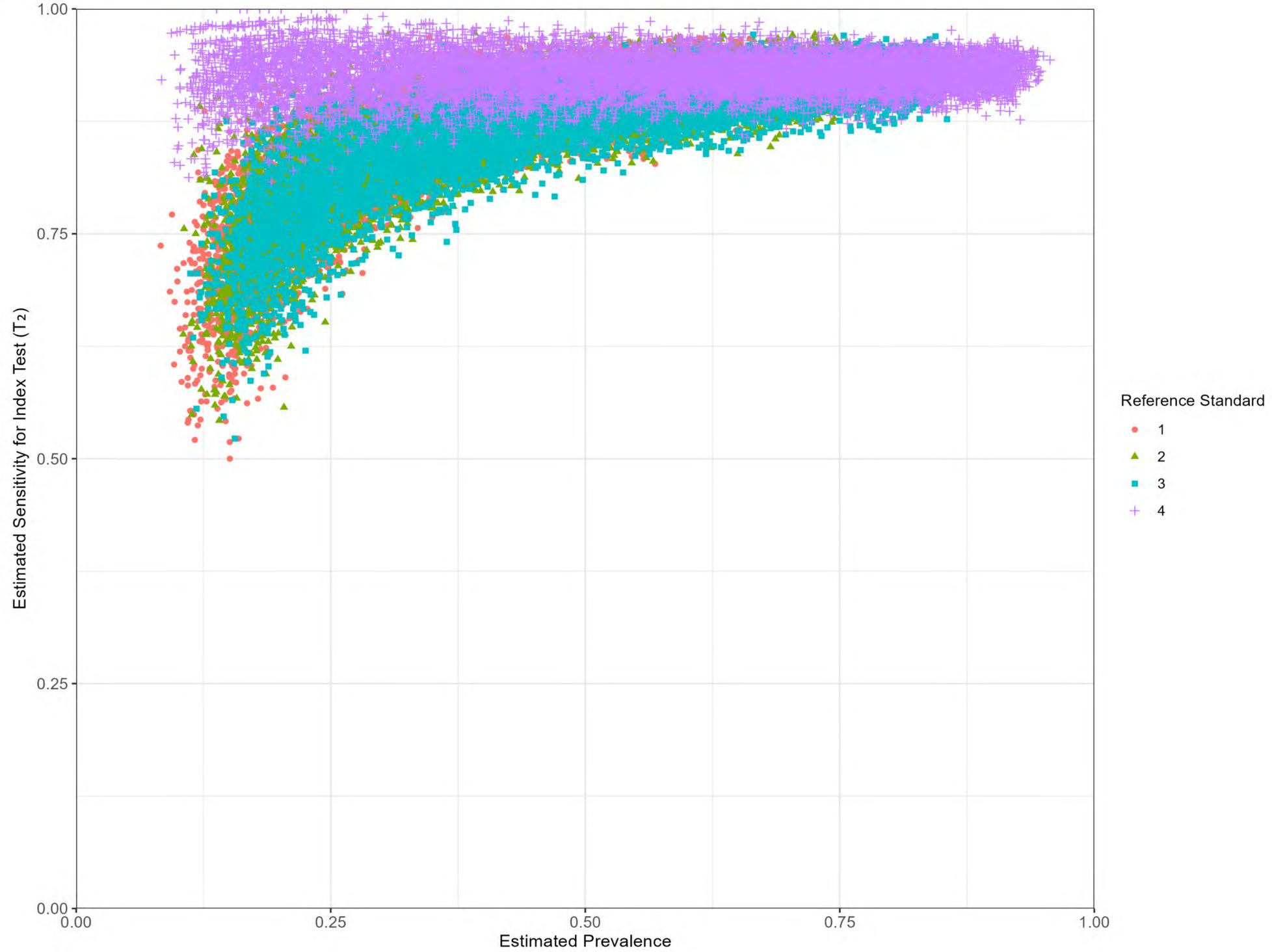
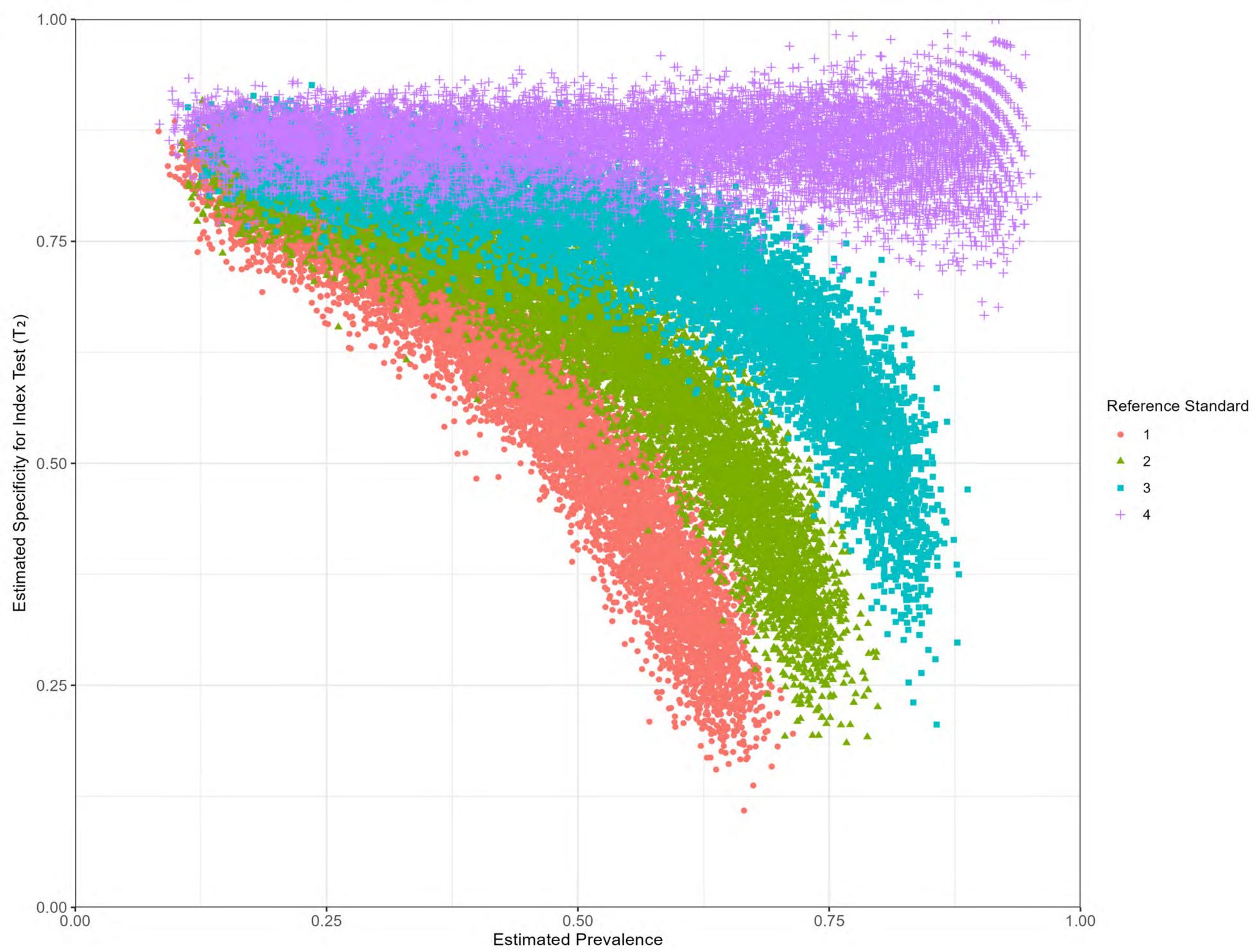

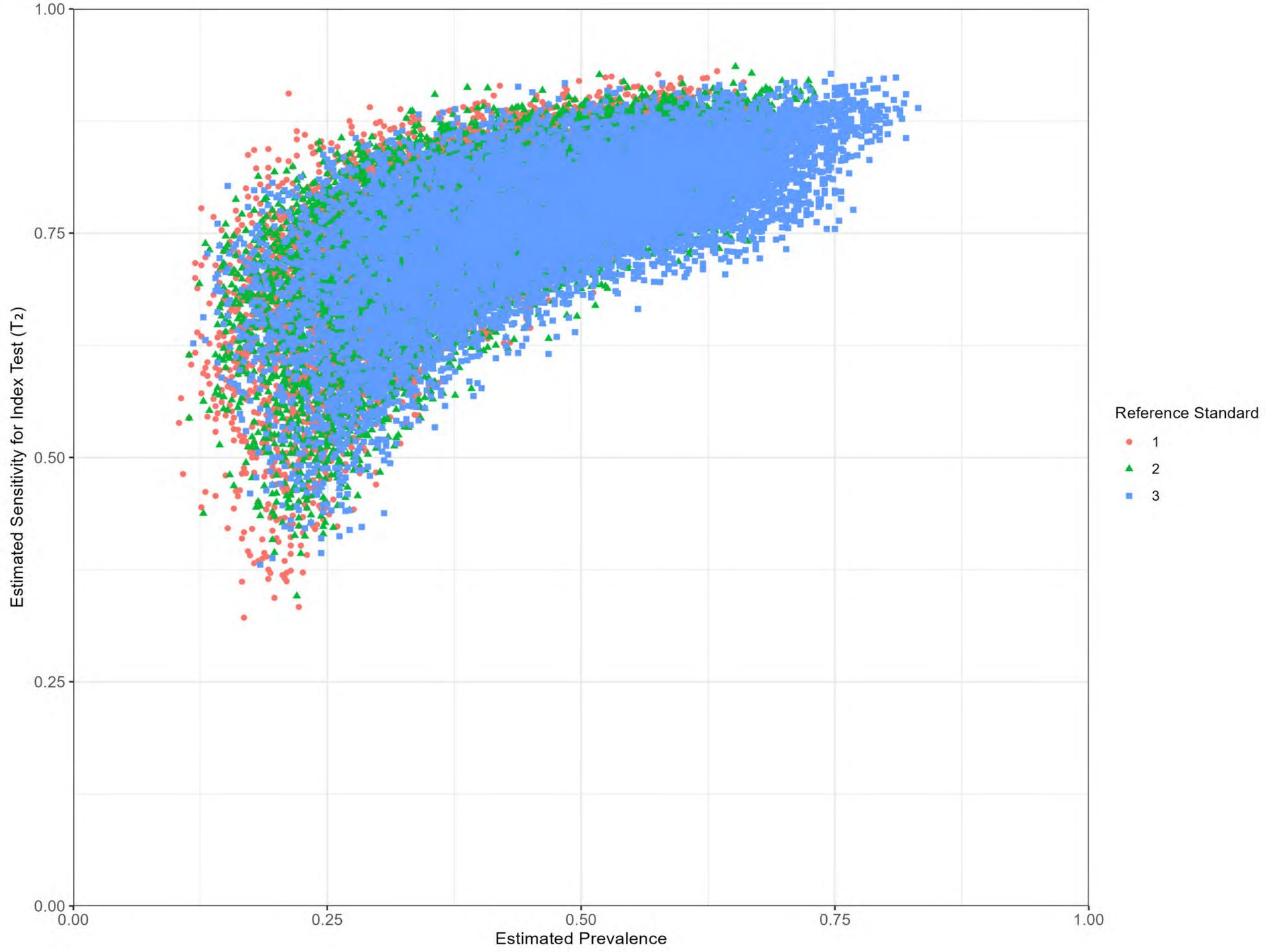
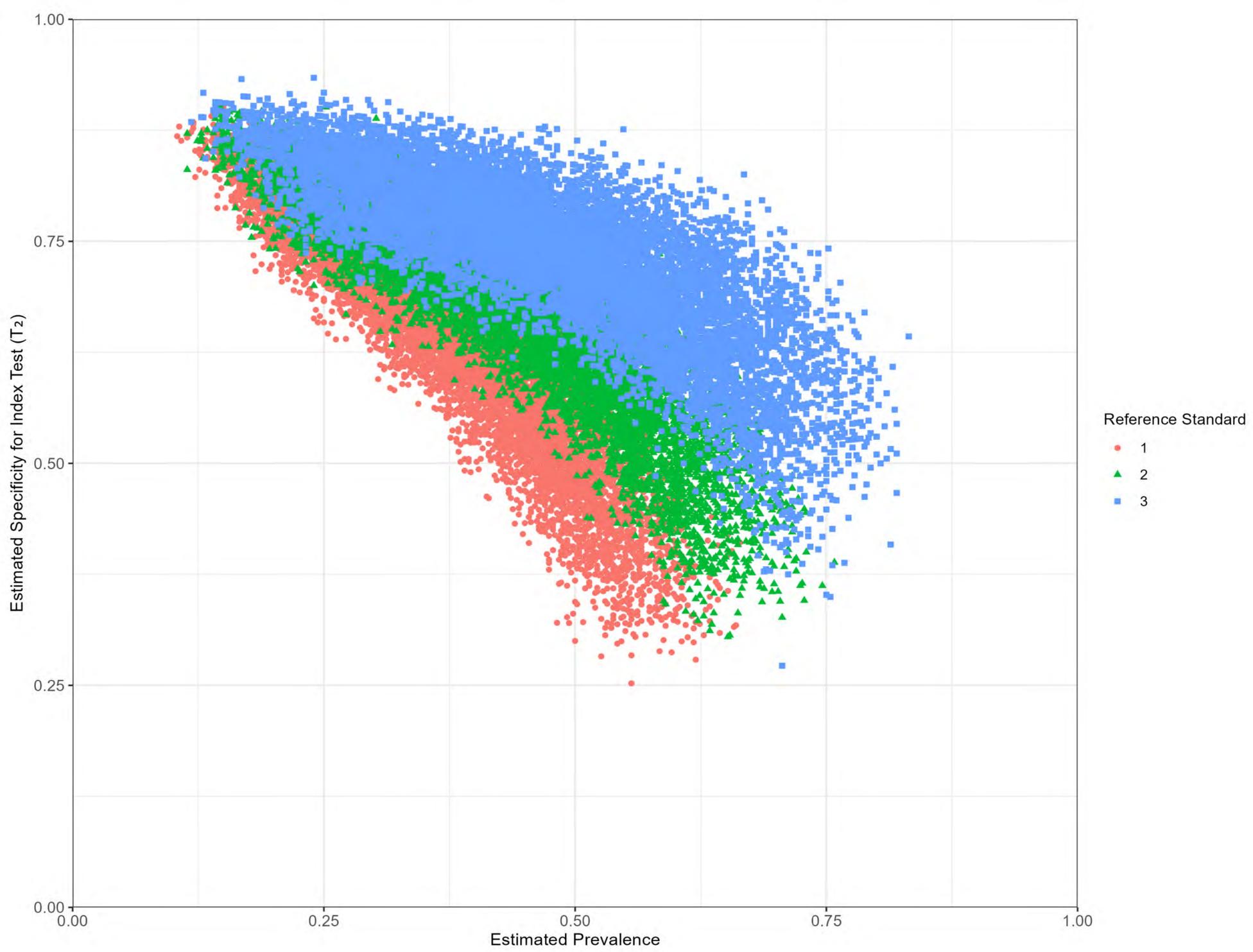

# Supplementary Materials

August 7, 2025

# Contents

# S1 - Simulation Design

In this section, we report the simulation design according to the ADEMP framework [1].

## S1.1 Aims of the Simulation

To examine the impact of five types of bias on the relationship between prevalence and index test accuracy estimates:

1. Reference standard error bias
2. Spectrum Effect
3. Confounding
4. Partial verification bias
5. Misassumption of conditional independence

## S1.2 Data-generating Mechanism

For each structure of bias, we consider 3 to 4 reference standard setups. For reference standard each setup, 10000 studies are simulated. In each simulated study, index test and reference standard results are simulated on 500 subjects.

The data generation process varies by bias structure but generally follows these steps:

1. Generate study-level prevalence for auxiliary covariate when applicable ($R$ in spectrum effect, confounding, and misassumption of conditional independence; $V$ in partial verification bias).

2. Generate study-level parameters: target condition prevalence, reference test accuracy and index test accuracy based on the bias structure.

3. For each subject, generate target condition $D$ based on study prevalence.

4. Generate reference standard and index test results for subjects conditional on $D$ and bias structure.

5. Obtain estimates of prevalence, index test sensitivity and index test specificity using reference standard assuming it is perfect.

We will use the values in Table S1 for our data-generating mechanism. Note that this procedure generates data at the subject-level, calculates accuracy and prevalence estimates for each of the 10,000 studies and examines associations across studies to demonstrate how subject-level biases can emerge as study-level patterns.

## S1.3 Estimand and other targets

For this simulation, we do not have an explicit estimand that appears in the data-generating mechanism. The primary target of this simulation is examining the association between the index test accuracy and the target condition prevalence estimates quantified using the Spearman rank correlation coefficient [2]. The secondary target is to apply latent class analysis model to account for each structure of bias to remove the association.

| Bias Structure | Data-generation Parameters |
|---|---|
| Reference Standard Error Bias | $D_{ij}|\pi_i \sim \text{Bern}(\pi_i)$; $\pi_i \sim \text{Unif}(0.1, 0.9)$; $T_{ijk}|D_{ij} = 1 \sim \text{Bern}(S_k)$; $T_{ijk}|D_{ij} = 0 \sim \text{Bern}(1 - C_k)$; $S_1 = 0.7, 0.8, 0.9, 1$; $C_1 = 0.95, 1$; $S_2 = 0.9$; $C_2 = 0.9$; $i = 1, 2, \ldots, 10000$; $j = 1, 2, \ldots, 500$; $k = 1, 2$ |
| Spectrum Effect | $D_{ij}|\pi_i \sim \text{Bern}(\pi_i)$; $\pi_i \sim \text{Unif}(0.1, 0.9)$; $T_{ijk}|D_{ij} = 1, R_{ij} = 1 \sim \text{Bern}(S_{k1})$; $T_{ijk}|D_{ij} = 1, R_{ij} = 0 \sim \text{Bern}(S_{k0})$; $T_{ijk}|D_{ij} = 0, R_{ij} = 1 \sim \text{Bern}(1 - C_{k1})$; $T_{ijk}|D_{ij} = 0, R_{ij} = 0 \sim \text{Bern}(1 - C_{k0})$; $R_{ij} \sim \text{Bern}(p_i)$; $p_i \sim \text{Beta}(1, 1)$; $S_{11} = S_{10} = 0.7, 0.8, 0.9, 1$; $C_{11} = C_{10} = 0.95, 1$; $S_{21} = 0.8$; $S_{20} = 0.9$; $C_{21} = 0.8$; $C_{20} = 0.9$; $i = 1, 2, \ldots, 10000$; $j = 1, 2, \ldots, 500$; $k = 1, 2$ |
| Confounding | $D_{ij}|\pi_i \sim \text{Bern}(\pi_i)$; $\pi_i|R_{ij} = 1 \sim \text{Unif}(0.7, 0.9)$; $\pi_i|R_{ij} = 0 \sim \text{Unif}(0.1, 0.3)$; $R_{ij} \sim \text{Bern}(p_i)$; $p_i \sim \text{Beta}(1, 1)$; $T_{ijk}|D_{ij} = 1, R_{ij} = 1 \sim \text{Bern}(S_{k1})$; $T_{ijk}|D_{ij} = 1, R_{ij} = 0 \sim \text{Bern}(S_{k0})$; $T_{ijk}|D_{ij} = 0, R_{ij} = 1 \sim \text{Bern}(1 - C_{k1})$; $T_{ijk}|D_{ij} = 0, R_{ij} = 0 \sim \text{Bern}(1 - C_{k0})$; $S_{11} = S_{10} = 0.7, 0.8, 0.9, 1$; $C_{11} = C_{10} = 0.95, 1$; $S_{21} = 0.8$; $S_{20} = 0.9$; $C_{21} = 0.8$; $C_{20} = 0.9$; $i = 1, 2, \ldots, 10000$; $j = 1, 2, \ldots, 500$; $k = 1, 2$ |
| Partial Verification Bias | $D_{ij}|\pi_i \sim \text{Bern}(\pi_i)$; $\pi_i \sim \text{Unif}(0.1, 0.9)$; $V_{ij} = 1|T_{ij2} = 1 \sim \text{Bern}(1)$; $V_{ij} = 1|T_{ij2} = 0, pv_i \sim \text{Bern}(pv_i)$; $pv_i \sim \text{Uniform}(0.5, 0.9)$; $T_{ijk}|D_{ij} = 1 \sim \text{Bern}(S_k)$; $T_{ijk}|D_{ij} = 0 \sim \text{Bern}(1 - C_k)$; $S_1 = 0.7, 0.8, 0.9, 1$; $C_1 = 0.95, 1$; $S_2 = 0.9$; $C_2 = 0.9$; $i = 1, 2, \ldots, 10000$; $j = 1, 2, \ldots, 500$; $k = 1, 2$ |
| Bias due to Misassumption of Conditional Independence | $D_{ij}|\pi_i \sim \text{Bern}(\pi_i)$; $\pi_i \sim \text{Unif}(0.1, 0.9)$; $R_{ij} \sim \text{Bern}(p_i)$; $p_i \sim \text{Beta}(1, 1)$; $T_{ijk}|D_{ij} = 1, R_{ij} = 1 \sim \text{Bern}(S_{k1})$; $T_{ijk}|D_{ij} = 0, R_{ij} = 1 \sim \text{Bern}(1 - C_{k1})$; $T_{ijk}|D_{ij} = 1, R_{ij} = 0 \sim \text{Bern}(S_{k0})$; $T_{ijk}|D_{ij} = 0, R_{ij} = 0 \sim \text{Bern}(1 - C_{k0})$; $S_{11} = 0.6, 0.7, 0.8$; $C_{11} = 0.85$; $S_{10} = 0.7, 0.8, 0.9$; $C_{10} = 0.95$; $S_{21} = 0.8$; $C_{21} = 0.8$; $S_{20} = 0.9$; $C_{20} = 0.9$; $i = 1, 2, \ldots, 10000$; $j = 1, 2, \ldots, 500$; $k = 1, 2$ |

Table S1: Simulation Parameters for Each Bias Structure. $D$ represents the target condition, $T_{ij1}$ and $T_{ij2}$ represent the reference standard and index test results respectively for subject $j$ in study $i$, $R$ represents auxiliary covariates, $S$ and $C$ represent sensitivity and specificity parameters, and $V$ indicates verification status.

### S1.4 Methods

In each simulated study, the index test accuracy is evaluated against the reference standard, and the target condition prevalence is estimated using reference standard results, assuming the reference standard is perfect. To be specific,

- The sensitivity of the index test is estimated as the proportion of positives captured by the index test out of the positives identified by the reference standard.

- The specificity of the index test is estimated as the proportion of negatives captured by the index test out of the negatives identified by the reference standard.

- The prevalence of the target condition is estimated using the proportion of positives identified by the reference standard out of the total number of subjects.

- The Spearman correlation coefficient is calculated separately for two associations: first,

  between sensitivity and prevalence; and second, between specificity and prevalence.

- The scatter plot of index test accuracy estimates against prevalence estimates is used to visualize the associations.

Genuine

## S1.5   Performance measures

Since we do not have an explicit estimand in the data-generating mechanism that we wish to recover, the performance measure is not applicable to this simulation study.

# S2 - Latent Class Bivariate Model

## S2.1 Model Description

Let $N_{ixy}$ denote the number of subjects in study i with results x (0 = negative, 1 = positive) on the index test and y on the reference standard. It follows that

$$N_{ixy} \sim Multinomial(n_i, p_{ixy}) \qquad (1)$$

with

$$p_{ixy} = \pi_i[S_{i1}^x(1-S_{i1})^{1-x}S_{i2}^y(1-S_{i2})^{1-y}] + (1-\pi_i)[C_{i1}^{1-x}(1-C_{i1})^x C_{i2}^{1-y}(1-C_{i2})^y], \qquad (2)$$

where $\pi_i$ denotes the prevalence of the target condition in study $i$, $S_{i1}$ and $C_{i1}$ are the $S$ and $C$ of the reference standard, respectively, $S_{i2}$ and $C_{i2}$ are the $S$ and $C$ of the index test, respectively. Let $\theta_{S1} = g(S_{i1})$, $\theta_{C1} = g(C_{i1})$, $\theta_{S2} = g(S_{i2})$ and $\theta_{C2} = g(C_{i2})$, where $g(\cdot)$ is the logit transformation. The bivariate normal hierarchical prior on the accuracy of each test can be expressed as

$$\begin{pmatrix} \theta_{Sk} \\ \theta_{Ck} \end{pmatrix} \sim N(\mu_k, \Sigma_k) \qquad (3)$$

with

$$\mu_k = (\mu_{Sk}, \mu_{Ck})^T, \Sigma_k = \begin{pmatrix} \sigma_{Sk}^2 & \rho_k \sigma_{Sk} \sigma_{Ck} \\ \rho_k \sigma_{Sk} \sigma_{Ck} & \sigma_{Ck}^2 \end{pmatrix}, k = 1, 2 \qquad (4)$$

Weak, low-information prior distributions were used for all parameters. Specifically, we used $Beta(1,1)$ for $\pi_i$, $N(0, precision = 0.5)$ for $\mu_{Sk}$ and $\mu_{Ck}$, respectively, $Uniform(-1,1)$ for $\rho_k$, t-distribution centered and truncated above 0 with a scale of 16 and 1 degree of freedom (i.e., half-Cauchy prior distribution) for $\sigma_{Sk}$ and $\sigma_{Ck}$, respectively [3].

Bayesian inference using Monte Carlo Markov chain methods (MCMC) was implemented using the rjags package in the R software environment [4,5]. Three independent MCMC chains of 50000 iterations were run, and the first 25000 iterations were discarded from each chain as burn-in iterations. The convergence of MCMC chains was assessed by checking the potential scale reduction factor ($\hat{R}$) proposed by Gelman and Rubin [6] as well as the posterior trace plots and density plots for each parameter. The posterior distributions of the parameters of interest were summarized and reported using posterior quantiles. A corresponding JAGS implementation can be found in the next subsection.

## S2.2 JAGS Implementation

```
model {
#============#
# LIKELIHOOD #
#============#
for(i in 1:N) {
        # The 4 cells in the two-by-two table of the index test vs. the
            imperfect reference test in each study in the meta-analysis

        cell[i,1:4] ~ dmulti(prob[i,1:4], n[i])

        # Multinomial probabilities of the 4 cells of the two-by-two
            table expressed in terms of the disease prevalence (prev),
            the sensitivity (se1, se2) and specificity (sp1, sp2) of the
            two tests.
        prob[i,1] <- prev[i]*( se[i]     * se2[i]     ) + (1-prev[i])*(
            (1-sp[i])    * (1-sp2[i]))
        prob[i,2] <- prev[i]*( se[i]     * (1-se2[i]) ) + (1-prev[i])*(
            (1-sp[i])    * sp2[i]      )
        prob[i,3] <- prev[i]*( (1-se[i]) * se2[i]     ) + (1-prev[i])*(
            sp[i]        * (1-sp2[i]))
        prob[i,4] <- prev[i]*( (1-se[i]) * (1-se2[i]) ) + (1-prev[i])*(
            sp[i]        * sp2[i]      )

        # Prior Distribution on Prevalence
        prev[i] ~ dbeta(1,1)
}

#====================#
# HIERARCHICAL PRIORS #
#====================#
for(j in 1:N) {
        # Hierarchical Prior for Index Test
        logit(se[j])    <- l[j,1]
        logit(sp[j])    <- l[j,2]
        l[j,1:2] ~ dmnorm(mu[1:2], T[1:2,1:2])

        # Hierarchical Prior for Reference Standard
        logit(se2[j])   <- l2[j,1]
        logit(sp2[j])   <- l2[j,2]
        l2[j,1:2] ~ dmnorm(mu2[1:2], T2[1:2,1:2])

}

#==========================#
# HYPER PRIOR DISTRIBUTIONS #
#==========================#

#==================================================================
##                   Index Test
#==================================================================
### Prior for the Logit Transformed Sensitivity (mu[1]) and Specificity
    (mu[2])
mu[1] ~ dnorm(0,0.5)
mu[2] ~ dnorm(0,0.5)
T[1:2,1:2]<-inverse(SIGMA[1:2,1:2])

```

```
46  ### Between-study Variance-Covariance Matrix of Index Test
47  SIGMA[1,1] <- sigma[1]*sigma[1]
48  SIGMA[2,2] <- sigma[2]*sigma[2]
49  SIGMA[1,2] <- rho*sigma[1]*sigma[2]
50  SIGMA[2,1] <- rho*sigma[1]*sigma[2]
51  rho ~ dunif(-1,1) # or set rho = 0
52  sigma[1] ~ dt(0, 16, 1) T(0,)
53  sigma[2] ~ dt(0, 16, 1) T(0,)
54
55  ### Mean Sensitivity and Specificity of Index Test ####
56  Mean_Se <- 1/(1+exp(-mu[1]))
57  Mean_Sp <- 1/(1+exp(-mu[2]))
58
59  #====================================================================
60  ##                         Reference Standard
61  #====================================================================
62  ### Prior for the Logit Transformed Sensitivity (mu2[1]) and
        Specificity (mu2[2])
63  mu2[1] ~ dnorm(0,0.5)
64  mu2[2] ~ dnorm(0,0.5)
65  T2[1:2,1:2]<-inverse(SIGMA2[1:2,1:2])
66
67  #### Between-study Variance-Covariance Matrix of Reference Standard
68  SIGMA2[1,1] <- sigma2[1]*sigma2[1]
69  SIGMA2[2,2] <- sigma2[2]*sigma2[2]
70  SIGMA2[1,2] <- rho2*sigma2[1]*sigma2[2]
71  SIGMA2[2,1] <- rho2*sigma2[1]*sigma2[2]
72  rho2 ~ dunif(-1,1) # or set rho2 = 0
73  sigma2[1] ~ dt(0, 16, 1) T(0,)
74  sigma2[2] ~ dt(0, 16, 1) T(0,)
75
76  #### Mean Sensitivity and Specificity of Reference Standard ####
77  Mean_Se2<-1/(1+exp(-mu2[1]))
78  Mean_Sp2<-1/(1+exp(-mu2[2]))
79  }
```

# S3 - Latent Class Bivariate Model for Partial Verification Bias

## S3.1 Model Description

We employed a Bayesian approach that can simultaneously adjust for reference standard error bias and partial verification bias [7].

| Stage of the Design | | $T_1+$ | $T_1-$ |
|---|---|---|---|
| **Stage 1** | Total Subjects Tested on $T_1$ | $N_1$ | $N_0$ |
| | Subjects Verified by $T_2$ | $V_1$ | $V_0$ |
| | Subjects Not Verified by $T_2$ | $N_1 - V_1$ | $N_0 - V_0$ |
| **Stage 2** | $T_2+$ | $X_1$ | $X_0$ |
| | $T_2-$ | $V_1 - X_1$ | $V_0 - X_0$ |

Table S2: A Two-stage Partial Verification Design in Diagnostic Test Evaluation.

Table S2 illustrates the classic two-stage partial verification design in diagnostic test evaluation. For each study $i$ (with the index omitted for simplicity), let $N$ represent the number of subjects tested on the index test, $N_1$ represent the number of index test positive, $p_1$ represent the probability of index test positive, $\pi$ denote the target condition prevalence, $S_1$ denote the index test sensitivity, and $C_1$ denote the index test specificity. It follows that

Stage 1:

$$N_1 | p_1 \sim Binomial(N, p_1), \quad (5)$$
$$p_1 | \pi, S_1, C_1 = \pi S_1 + (1-\pi)(1-C_1). \quad (6)$$

Now, let $V_1$ and $V_0$ denote the number of subjects verified by reference standard after obtaining a positive and negative result on the initial index test, respectively. Let $X_1$ and $X_0$ represent the test positives with a probability of $q_1$ and $q_0$ out of $V_1$ and $V_0$, respectively. It follows that

Stage 2:

$$X_1 | q_1 \sim Binomial(V_1, q_1), \quad (7)$$
$$X_0 | q_0 \sim Binomial(V_0, q_0), \quad (8)$$
$$q_1 = \frac{\pi S_1 S_2 + (1-\pi)(1-C_1)(1-C_2)}{\pi S_1 + (1-\pi)(1-C_1)}, \quad (9)$$
$$q_0 = \frac{\pi(1-S_1)S_2 + (1-\pi)C_1(1-C_2)}{\pi(1-S_1) + (1-\pi)C_1}. \quad (10)$$

The pooling of $S_1$, $S_2$, $C_1$ and $C_2$ follows the same hierarchical structure as described in Subsection S2.1. Weak, low-information prior distributions were used for all parameters. Specifically, we used $Beta(1,1)$ for $\pi_i$, $N(0, precision = 0.01)$ for $\mu_{Sk}$ and $\mu_{Ck}$ (truncated above 0), respectively, $Uniform(0,2)$ for $\sigma_{Sk}$ and $\sigma_{Ck}$ to avoid singularity issue, respectively [3].

Bayesian inference using Monte Carlo Markov chain methods (MCMC) was implemented using the rjags package in the R software environment [4,5]. Three independent MCMC chains of 50000 iterations were run, and the first 25000 iterations were discarded from each chain as burn-in iterations. The convergence of MCMC chains was assessed by checking the potential scale reduction factor ($\hat{R}$) proposed by Gelman and Rubin [6] as well as the posterior trace plots and density plots for each parameter. The posterior distributions of the parameters of interest were summarized and reported using posterior quantiles. A corresponding meta-analysis implementation in JAGS can be found in the next subsection.

## S3.2 JAGS Implementation

```
model {
for(i in 1:num_studies){
    # observed: N1, N, X1, X0, V1, V0
    # Stage 1 #
    N1[i] ~ dbinom(p1[i], N[i])
    p1[i] <- prev[i] * se[i] + (1 - prev[i]) * (1 - sp[i])

    # Stage 2 #
    X1[i] ~ dbinom(q1[i], V1[i])
    X0[i] ~ dbinom(q0[i], V0[i])
    q1[i] <- (prev[i] * se[i] * se2[i] + (1-prev[i]) * (1-sp[i]) * (1-
        sp2[i]))/(prev[i] * se[i] + (1-prev[i]) * (1-sp[i]))
    q0[i] <- (prev[i] * (1-se[i]) * se2[i] + (1-prev[i]) * sp[i] * (1-
        sp2[i]))/(prev[i] * (1 - se[i]) + (1-prev[i]) * sp[i])

    # Prior Distributions #
    logit(se[i]) <- l1[i,1]
    logit(sp[i]) <- l1[i,2]
    l1[i,1:2] ~ dmnorm(mu1[1:2], T[1:2,1:2])

    logit(se2[i]) <- l2[i,1]
    logit(sp2[i]) <- l2[i,2]
    l2[i,1:2] ~ dmnorm(mu2[1:2], T2[1:2,1:2])

    prev[i] ~ dbeta(1, 1)
}

        # Hyperpriors for test 1
        mu1[1] ~ dnorm(0, 0.01)           # Mean logit sensitivity for
            test 1
        mu1[2] ~ dnorm(0, 0.01)T(0,)      # Mean logit specificity for
            test 1

        # Hyperpriors for test 2
        mu2[1] ~ dnorm(0, 0.01)           # Mean logit sensitivity for
            test 2
        mu2[2] ~ dnorm(0, 0.01)T(0,)      # Mean logit specificity for
            test 2

        #### BETWEEN-STUDY VARIANCE-COVARIANCE MATRIX
        T[1:2,1:2] <- inverse(SIGMA[1:2,1:2])
        SIGMA[1,1] <- sigma[1]*sigma[1]
        SIGMA[2,2] <- sigma[2]*sigma[2]
        SIGMA[1,2] <- rho*sigma[1]*sigma[2]
        SIGMA[2,1] <- rho*sigma[1]*sigma[2]

        T2[1:2,1:2] <- inverse(SIGMA2[1:2,1:2])
        SIGMA2[1,1] <- sigma2[1]*sigma2[1]
        SIGMA2[2,2] <- sigma2[2]*sigma2[2]
        SIGMA2[1,2] <- rho2*sigma2[1]*sigma2[2]
        SIGMA2[2,1] <- rho2*sigma2[1]*sigma2[2]

        sigma[1] ~ dunif(0, 2)      # SD for logit sensitivity of test 1
        sigma[2] ~ dunif(0, 2)      # SD for logit specificity of test 1
        rho = 0            # Correlation between logit sensitivity and
            specificity of test 1
```

```
            sigma2[1] ~ dunif(0, 2)      # SD for logit sensitivity of test 2
            sigma2[2] ~ dunif(0, 2)      # SD for logit specificity of test 2
            rho2 = 0            # Correlation between logit sensitivity and
                specificity of test 2

            # Other #
            mean_se  <- 1 / (1 + exp(-mu1[1]))
            mean_sp  <- 1 / (1 + exp(-mu1[2]))
            mean_se2 <- 1 / (1 + exp(-mu2[1]))
            mean_sp2 <- 1 / (1 + exp(-mu2[2]))
}
```

# S4 - Extended Scenarios of Partial Verification Bias Across Different Verification Rates

In this section, we present extended scenario results under partial verification bias with two distinct verification rates applied separately. These supplementary analyses build upon our main findings by examining how varying verification rates (0.7-0.9 or 0.1-0.9) affect the association between estimated accuracy and prevalence.

Figure S1 and Figure S2 show the results when a higher verification rate is employed. Spearman correlations with prevalence: sensitivity (0.842, 0.841, 0.830, and -0.017) and specificity (-0.973, -0.965, -0.930, and -0.012) for Setups 1-4, respectively.

Figure S3 and Figure S4 show the results when a lower verification rate is employed. Spearman correlations with prevalence: sensitivity (0.744, 0.717, 0.716, and 0.230) and specificity (-0.922, -0.898, -0.834, and -0.260) for Setups 1-4, respectively.

While sensitivity and prevalence estimates exhibited similar patterns across different verification rates, specificity estimates showed noticeable differences. Specifically, lower verification rates produced larger spread in specificity estimates and increased magnitude of association between specificity and prevalence estimates. These changes in association strength were also reflected in the correlation coefficients between specificity and prevalence estimates, regardless of whether a perfect or imperfect reference standard was employed.

## S4.1 High Verification Rate (0.7-0.9)

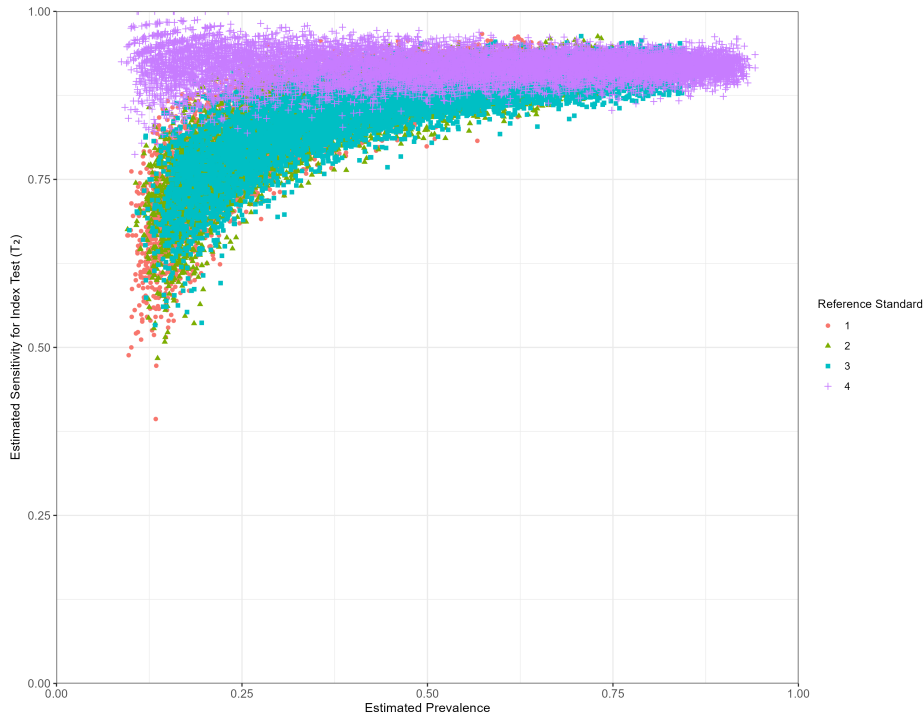

Figure S1: Relationship between estimated sensitivity of index test $T_2$ and estimated prevalence under partial verification bias (Verification Rate: 0.7-0.9).

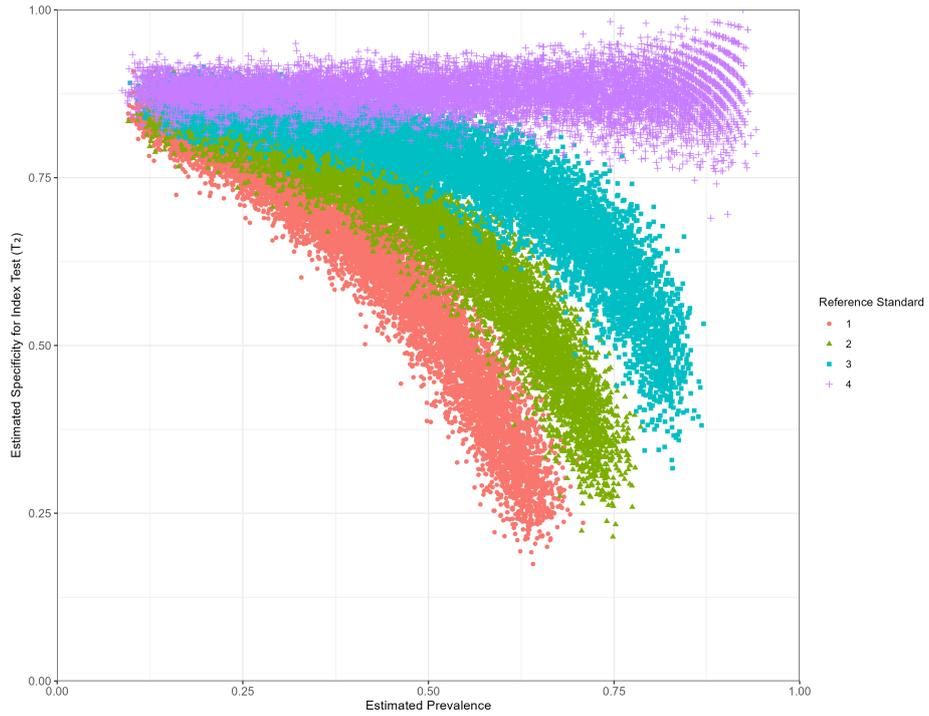

Figure S2: Relationship between estimated specificity of index test $T_2$ and estimated prevalence under partial verification bias (Verification Rate: 0.7-0.9).

### S4.2 Low Verification Rate (0.1-0.9)

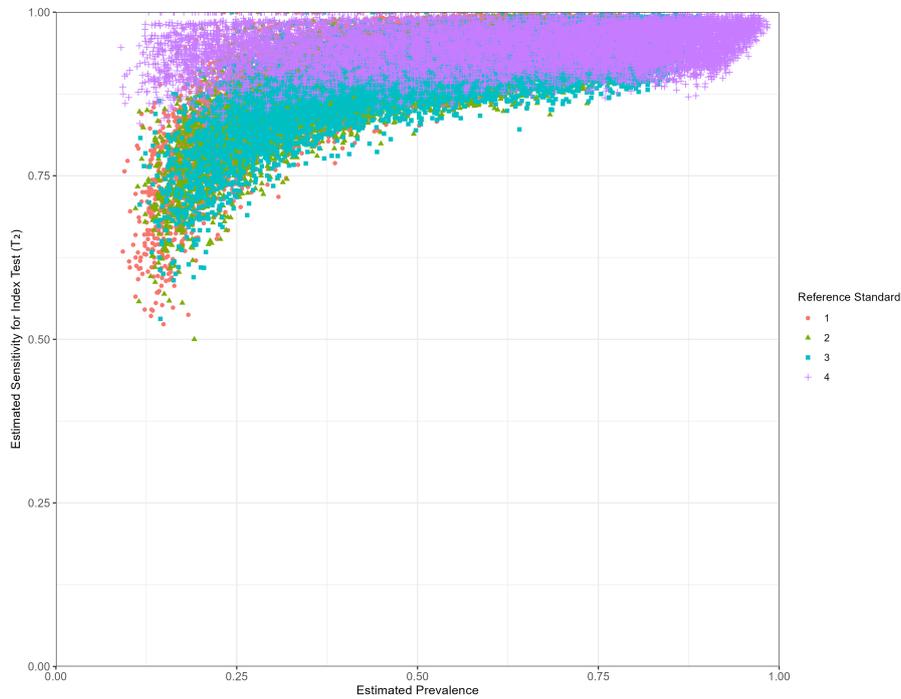

Figure S3: Relationship between estimated sensitivity of index test $T_2$ and estimated prevalence under partial verification bias (Verification Rate: 0.1-0.9).

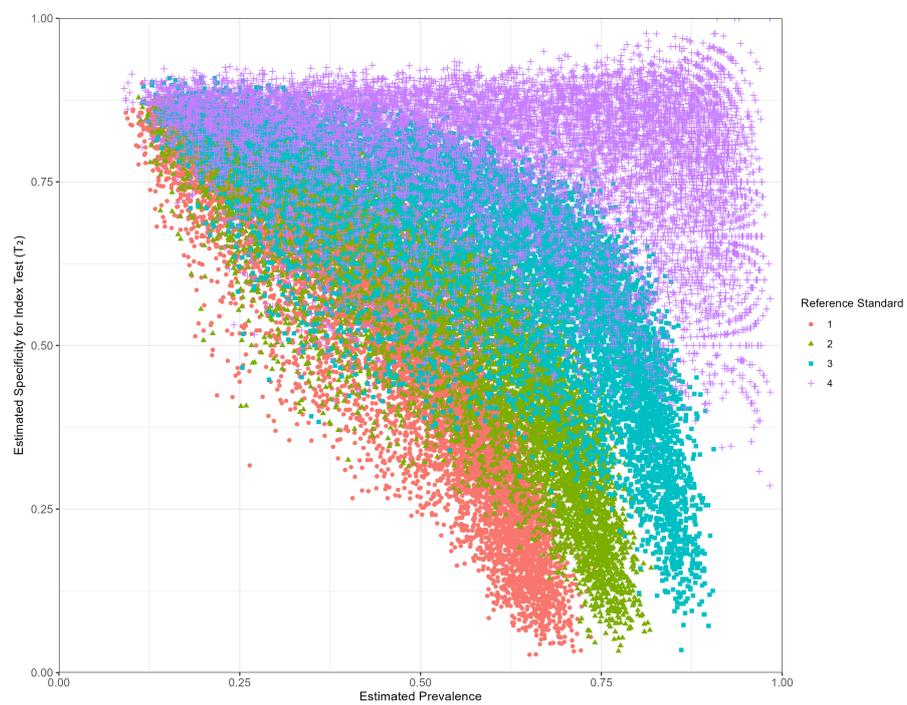

Figure S4: Relationship between estimated specificity of index test $T_2$ and estimated prevalence under partial verification bias (Verification Rate: 0.1-0.9).

# S5 - Adjusted Association Plot

## S5.1 Adjusted Association Plot - Reference Standard Error Bias

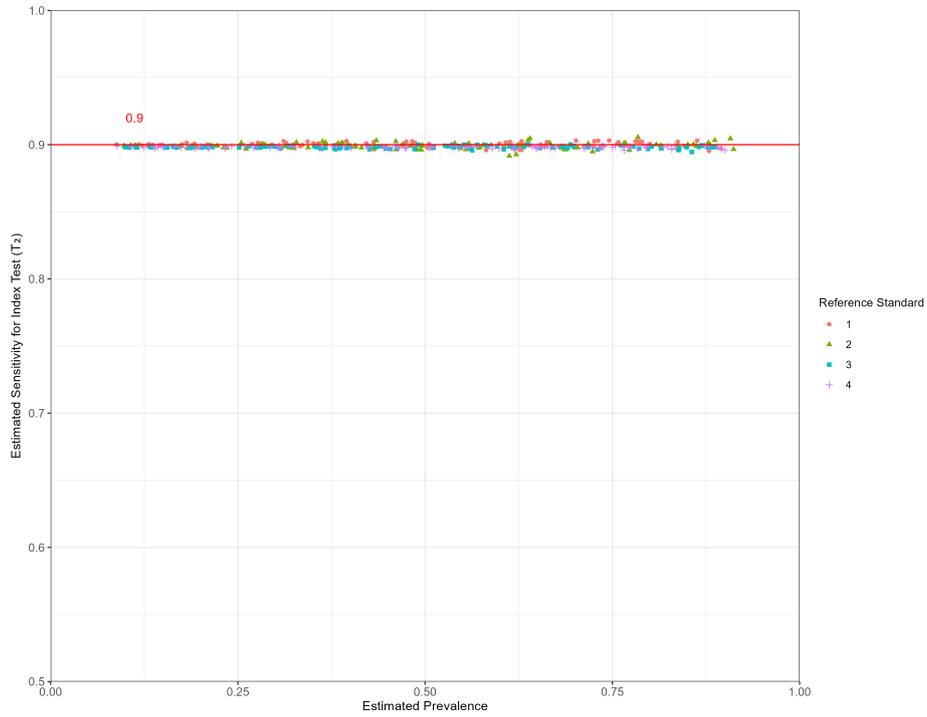

Figure S5: Relationship between estimated sensitivity of index test $T_2$ and estimated prevalence under reference standard error bias after latent class adjustment.

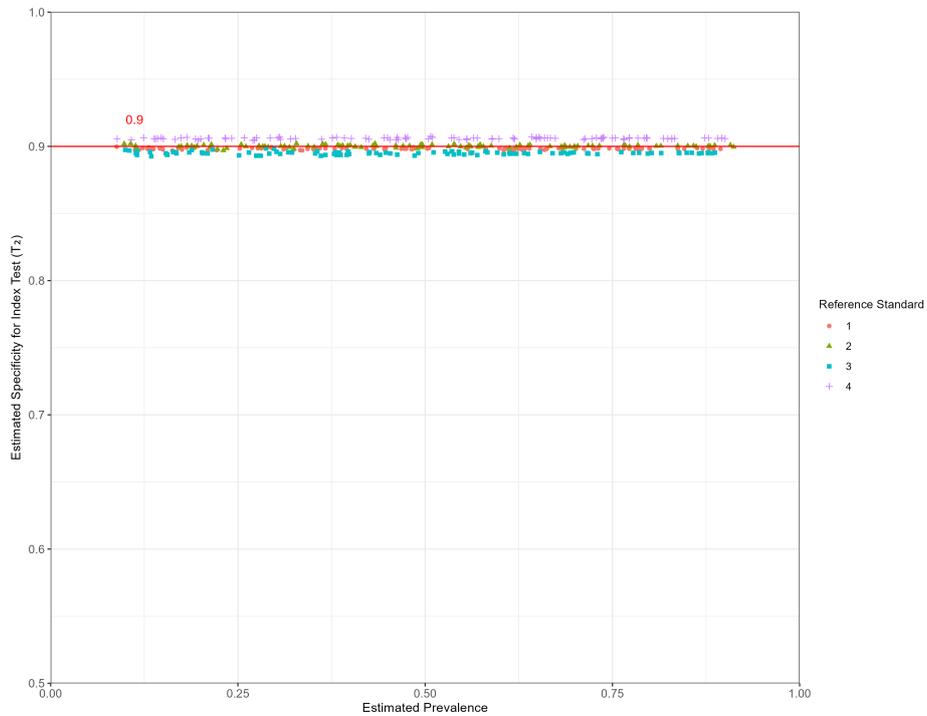

Figure S6: Relationship between estimated specificity of index test $T_2$ and estimated prevalence under reference standard error bias after latent class adjustment.

## S5.2 Adjusted Association Plot - Spectrum Effect

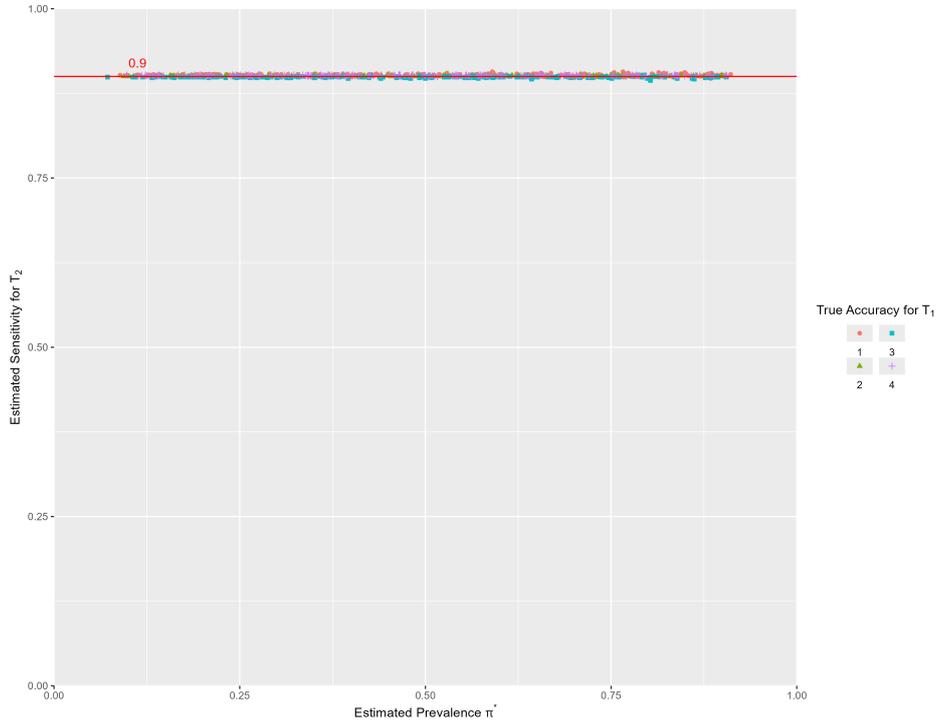

Figure S7: Relationship between estimated sensitivity of index test $T_2$ and estimated prevalence under spectrum effect after latent class adjustment in R negative group.

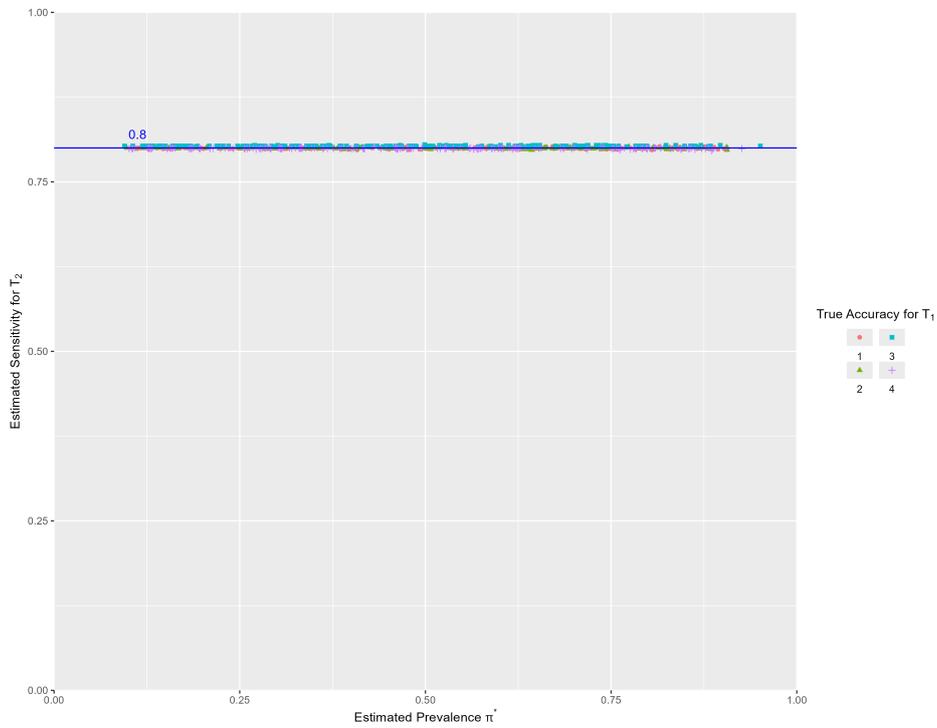

Figure S8: Relationship between estimated sensitivity of index test $T_2$ and estimated prevalence under spectrum effect after latent class adjustment in R positive group.

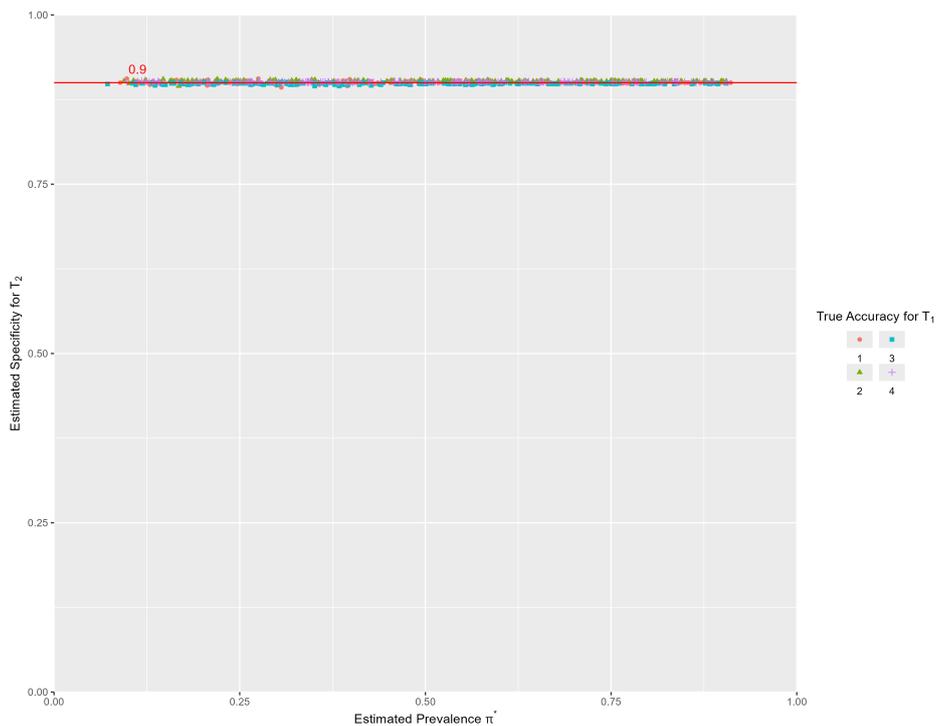

Figure S9: Relationship between estimated specificity of index test $T_2$ and estimated prevalence under spectrum effect after latent class adjustment in R negative group.

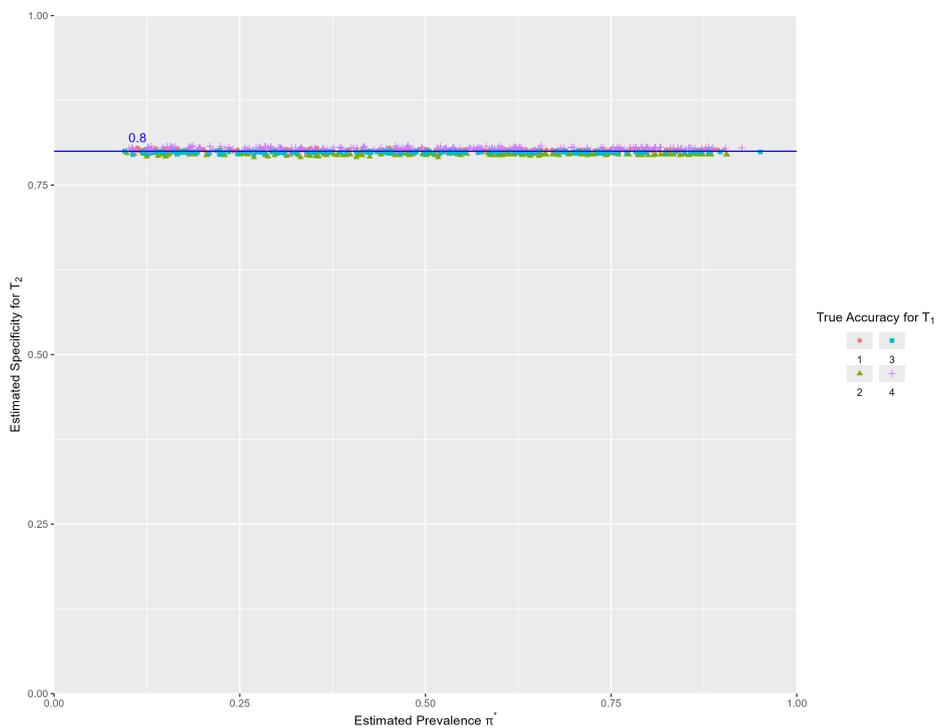

Figure S10: Relationship between estimated specificity of index test $T_2$ and estimated prevalence under spectrum effect after latent class adjustment in R positive group.

## S5.3 Adjusted Association Plot - Confounding

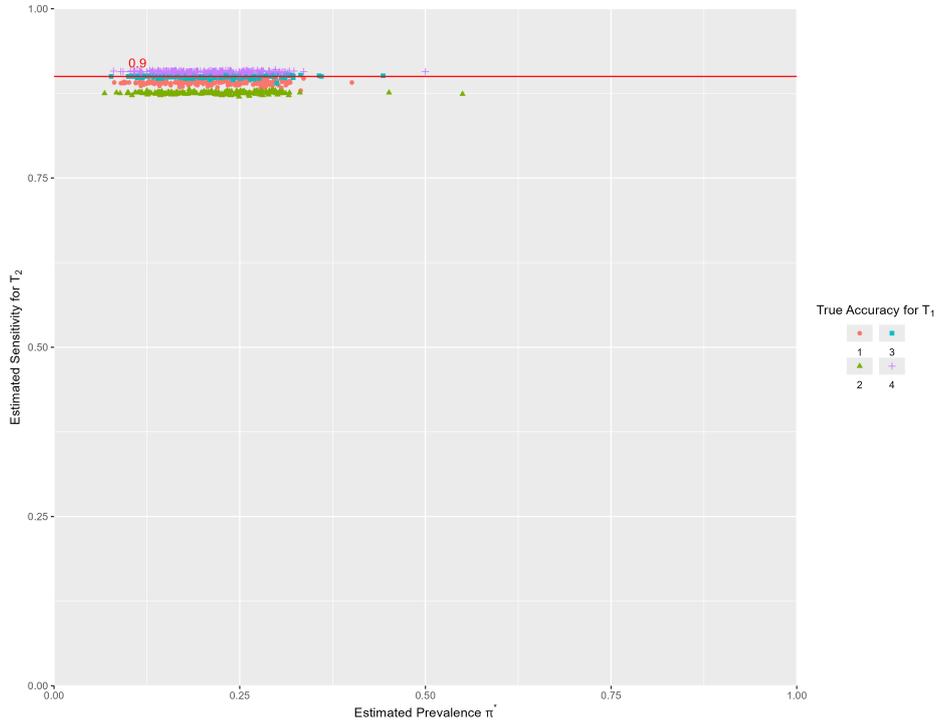

Figure S11: Relationship between estimated sensitivity of index test $T_2$ and estimated prevalence under confounding after latent class adjustment in R negative group.

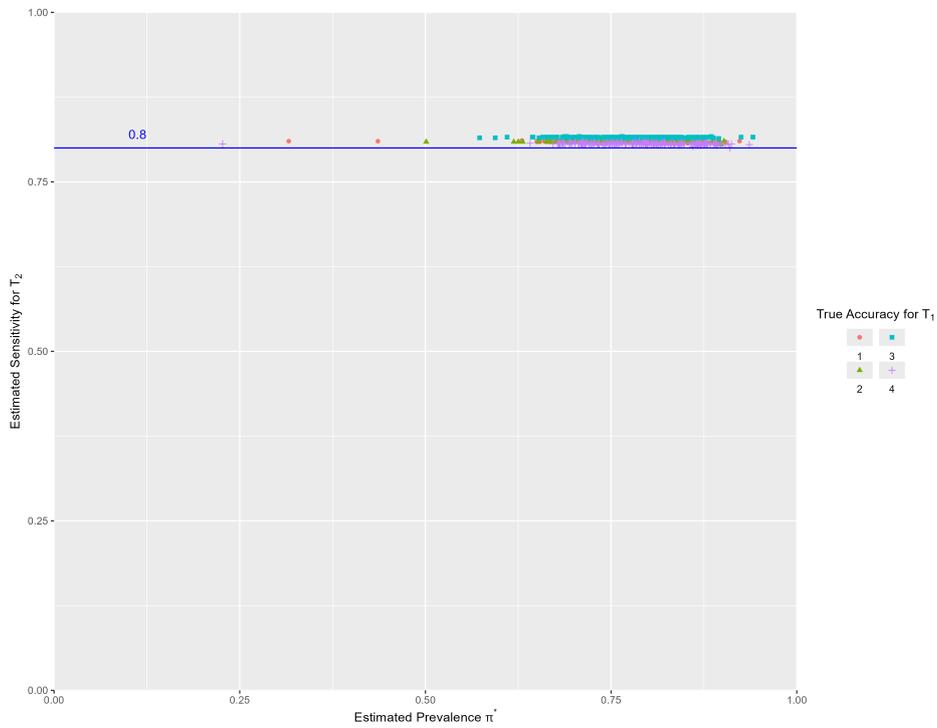

Figure S12: Relationship between estimated sensitivity of index test $T_2$ and estimated prevalence under confounding after latent class adjustment in R positive group.

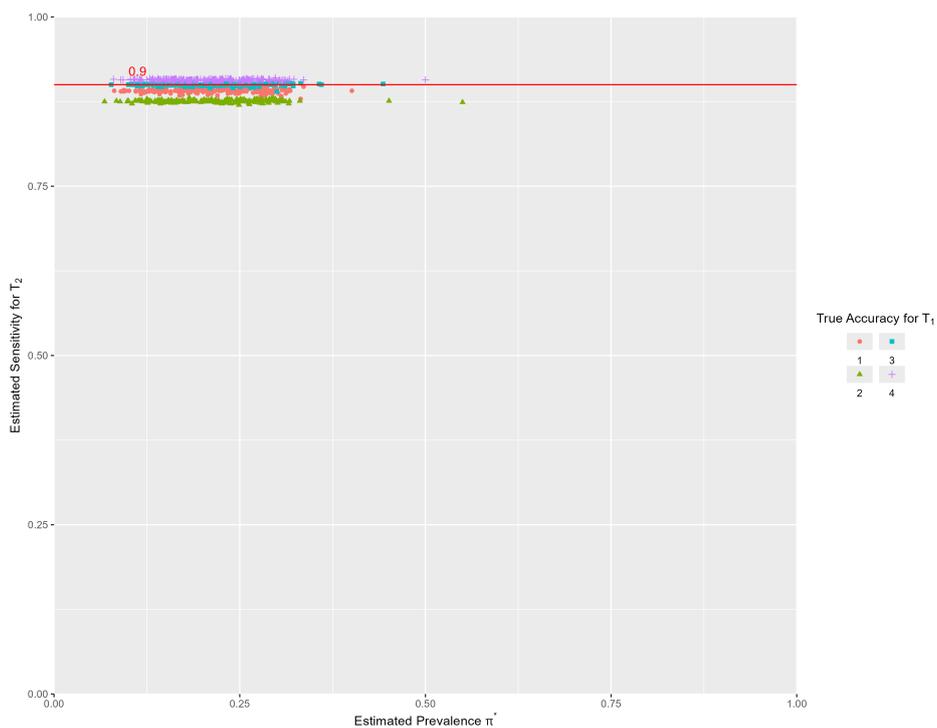

Figure S13: Relationship between estimated specificity of index test $T_2$ and estimated prevalence under confounding after latent class adjustment in R negative group.

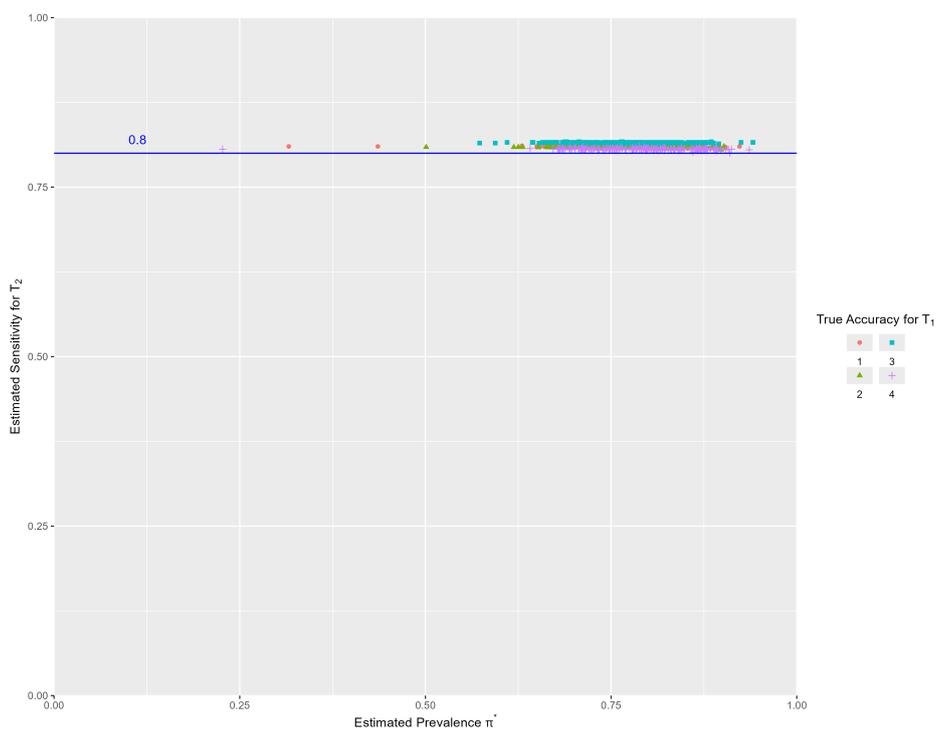

Figure S14: Relationship between estimated specificity of index test $T_2$ and estimated prevalence under confounding after latent class adjustment in R positive group.

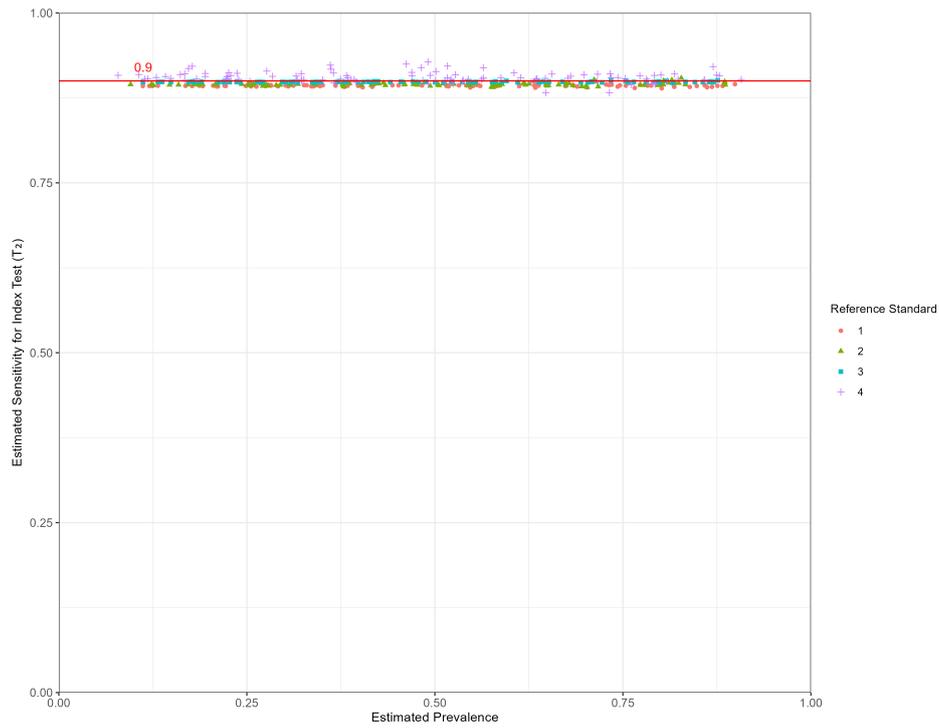

Figure S15: Relationship between estimated sensitivity of index test $T_2$ and estimated prevalence under partial verification bias after latent class adjustment.

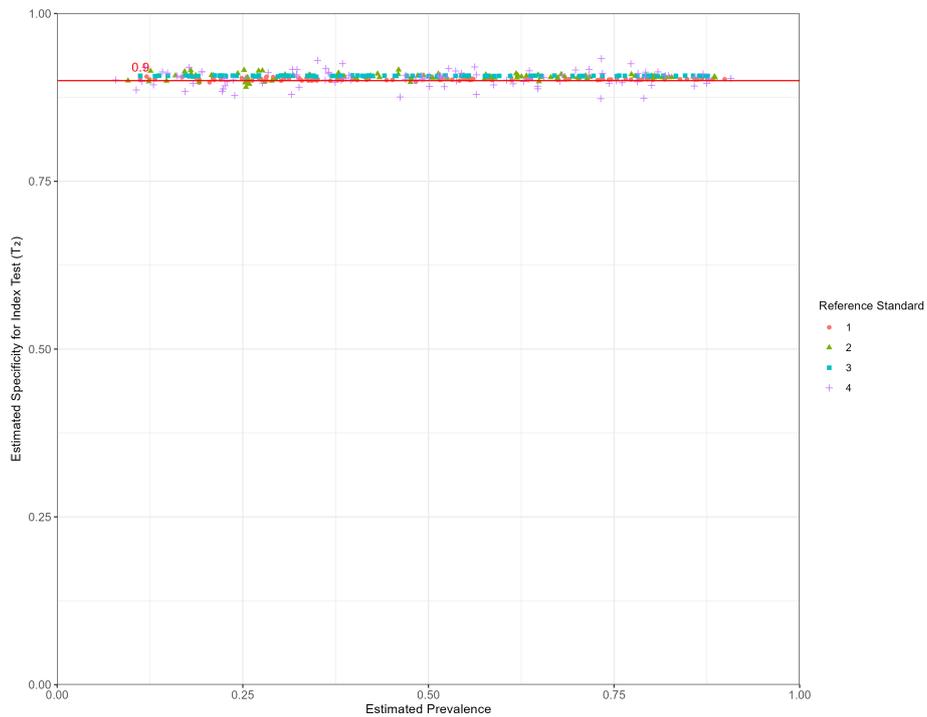

Figure S16: Relationship between estimated specificity of index test $T_2$ and estimated prevalence under partial verification bias bias after latent class adjustment.

## S5.4 Adjusted Association Plot - Partial Verification Bias

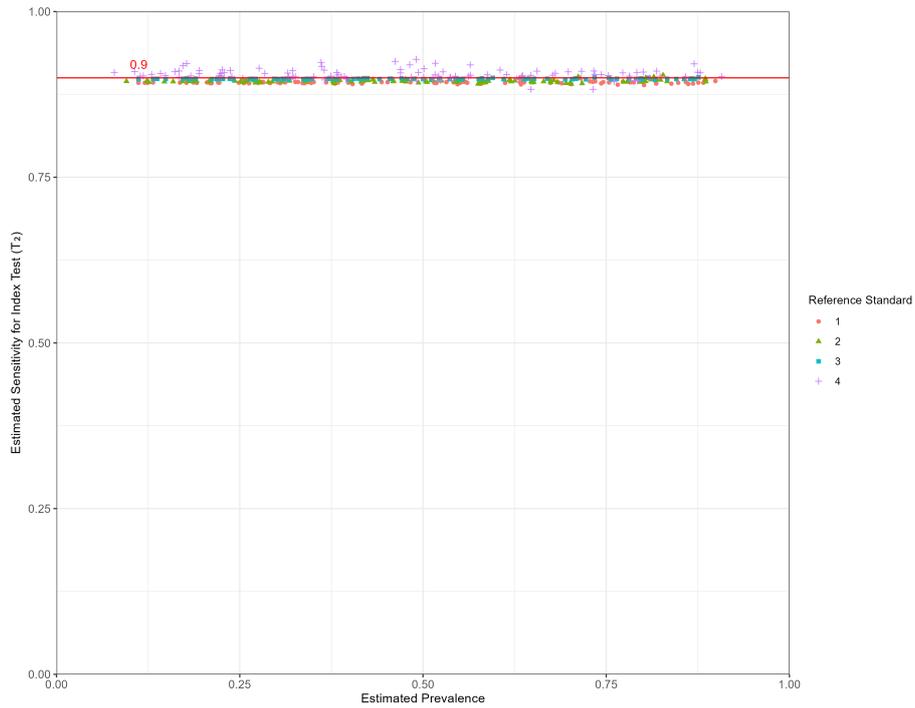

Figure S17: Relationship between estimated sensitivity of index test $T_2$ and estimated prevalence under partial verification bias after latent class adjustment in R negative group.

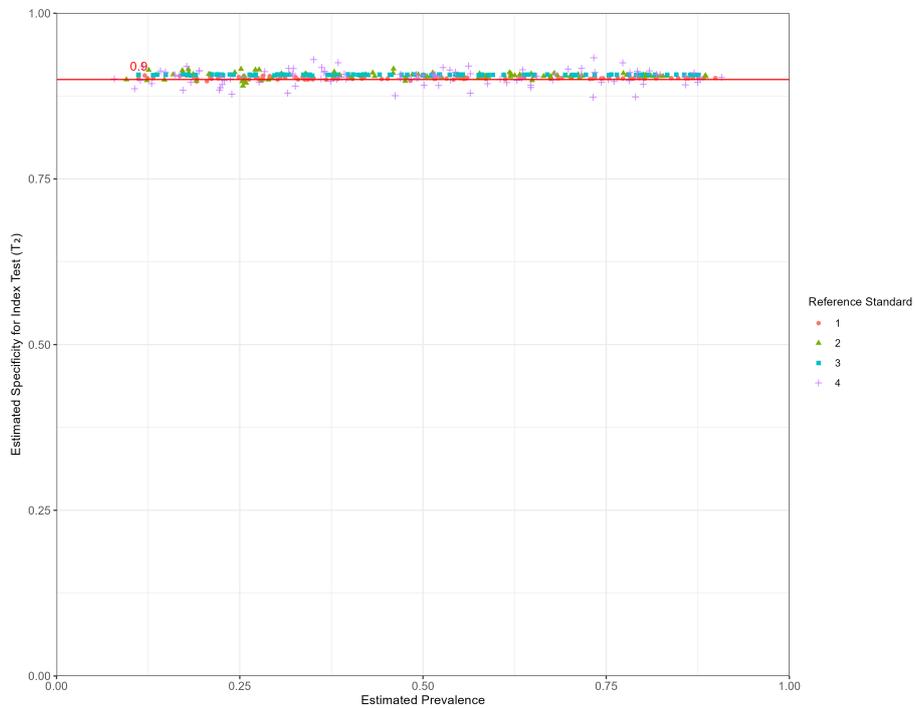

Figure S18: Relationship between estimated specificity of index test $T_2$ and estimated prevalence under partial verification bias after latent class adjustment in R negative group.

## S5.5 Adjusted Association Plot - Misassumption of Conditional Independence

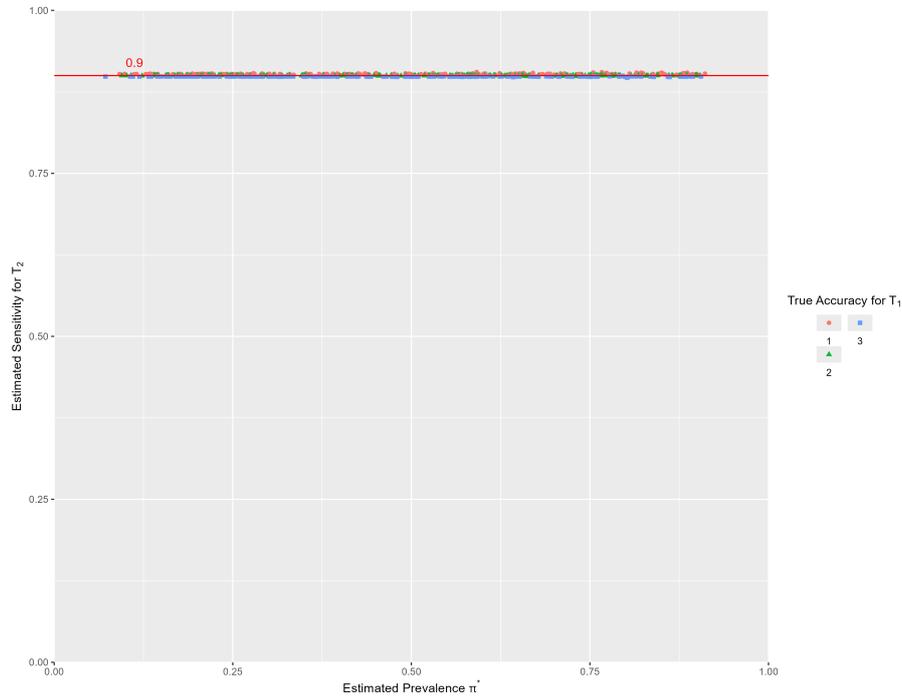

Figure S19: Relationship between estimated sensitivity of index test $T_2$ and estimated prevalence under misassumption of conditional dependence after latent class adjustment in R negative group.

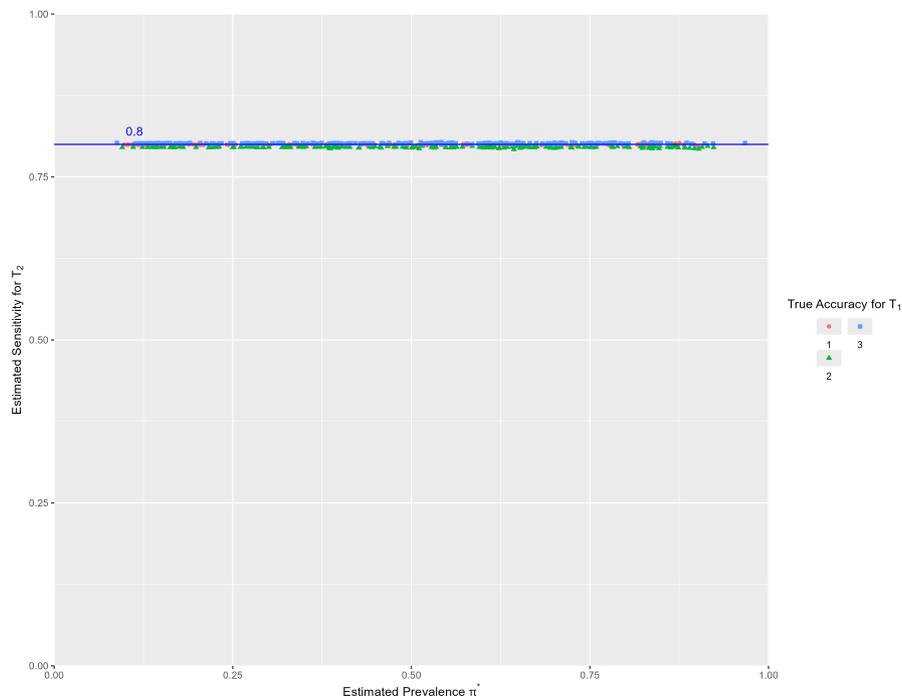

Figure S20: Relationship between estimated sensitivity of index test $T_2$ and estimated prevalence under misassumption of conditional dependence after latent class adjustment in R positive group.

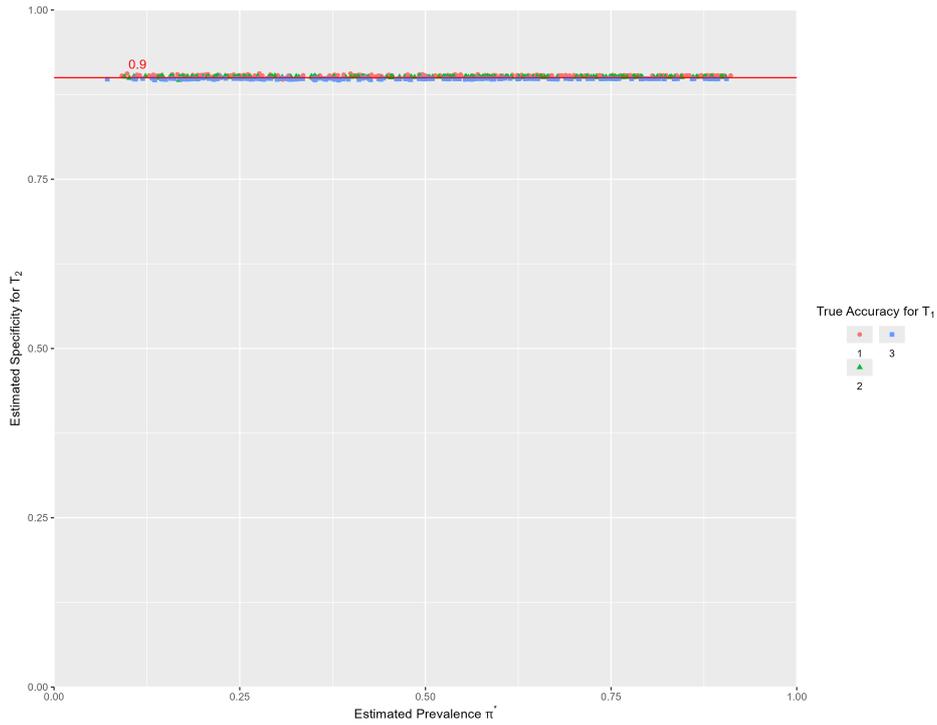

Figure S21: Relationship between estimated specificity of index test $T_2$ and estimated prevalence under misassumption of conditional dependence after latent class adjustment in R negative group.

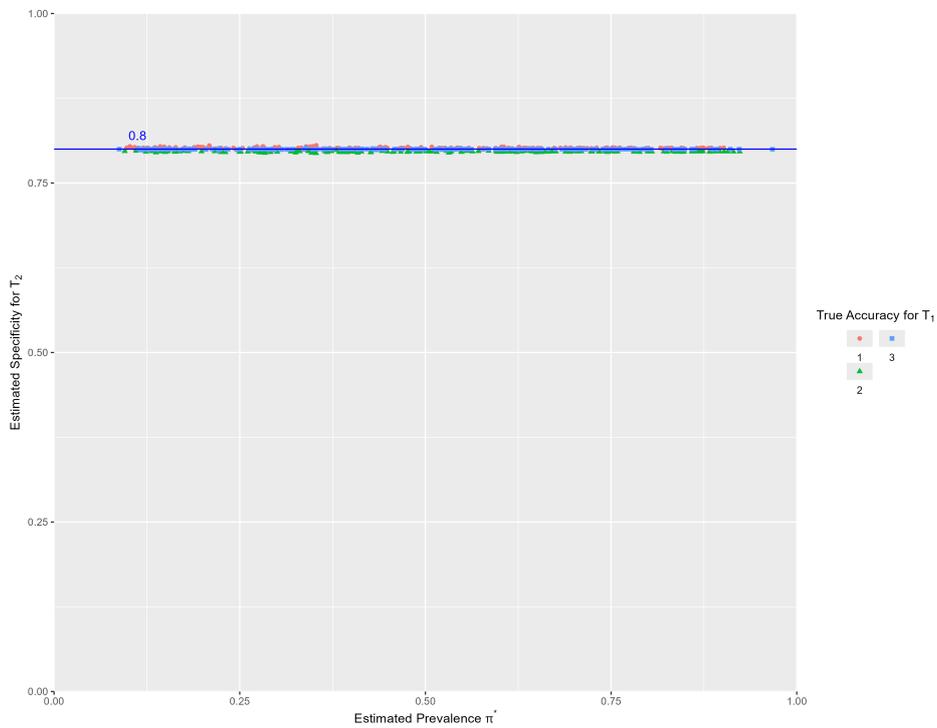

Figure S22: Relationship between estimated specificity of index test $T_2$ and estimated prevalence under misassumption of conditional dependence after latent class adjustment in R positive group.